\documentclass[lettersize,journal]{IEEEtran}
\usepackage{pifont}
\usepackage{array}
\usepackage{booktabs}
\usepackage{multirow}
\usepackage{subfigure}
\usepackage{times,amsmath,color,amssymb,amsfonts,graphicx,epsfig,cite,psfrag,subfigure,algorithm,epstopdf,bm}
\usepackage{amsmath}
\usepackage{algorithm}
\usepackage{algpseudocode}
\usepackage{textcomp}
\def\BibTeX{{\rm B\kern-.05em{\sc i\kern-.025em b}\kern-.08em
    T\kern-.1667em\lower.7ex\hbox{E}\kern-.125emX}}

\newcommand\eps\varepsilon

\begin{document}


\def\FIGDIR{./figs}

\IEEEoverridecommandlockouts

\title{Knowledge Sharing-enabled Semantic Rate Maximization for Multi-cell Task-oriented Hybrid Semantic-Bit Communication Networks}

\author{Hong Chen, \IEEEmembership{Member, IEEE}, Fang Fang, \IEEEmembership{Senior Member, IEEE}, and Xianbin Wang, \IEEEmembership{Fellow, IEEE}

\thanks{
Hong Chen, Fang Fang, and Xianbin Wang are with the Department of Electrical and Computer Engineering, and Fang Fang is also with the Department of Computer Science, Western University, London, ON N6A 3K7, Canada (e-mail: \{hche88, fang.fang, xianbin.wang\}@uwo.ca).

}
}

\maketitle \thispagestyle{empty}

\begin{abstract}
In task-oriented semantic communications, the transmitters are designed to deliver task-related semantic information rather than every signal bit to receivers, which alleviates the spectrum pressure by reducing network traffic loads. Effective semantic communications depend on the perfect alignment of shared knowledge between transmitters and receivers, however, the alignment of knowledge cannot always be guaranteed in practice. To tackle this challenge, we propose a novel knowledge sharing-enabled task-oriented hybrid semantic and bit communications mechanism, where a mobile device (MD) can proactively share and upload the task-related mismatched knowledge to associated small base station (SBS). The traditional bit communications can be adopted as an aid to transmit the rest data related to unshared mismatched knowledge to guarantee the effective execution of target tasks. Considering the heterogeneous transceivers in multi-cell networks, target task demands, and channel conditions, an optimization problem is formulated to maximize the generalized effective semantic transmission rate of all MDs by jointly optimizing knowledge sharing, semantic extraction ratio, and SBS association, while satisfying the semantic accuracy requirements and delay tolerances of MD target tasks. The formulated mixed integer nonlinear programming problem is decomposed into multiple subproblems equivalently. An optimum algorithm is proposed and another efficient algorithm is further developed using hierarchical class partitioning and monotonic optimization. Simulation results demonstrate the validity and superior performance of proposed solutions.
\end{abstract}


\sloppy
\allowdisplaybreaks

\vspace{-8mm}
\section{Introduction}
\label{sec:introduction}

\IEEEPARstart{S}{emantic} communications, as an emerging intelligent communication paradigm, are realized by exchanging the meaning  or semantics extracted from messages rather than transmitting the raw bit stream messages directly \cite{Future,SC,SAN}. By leveraging shared knowledge bases between the transmitter and receiver, more meaningful semantic information is extracted from the source data prior to undergoing semantic and channel encoding at the transmitter. The receiver can recover the source information by semantic restoration from the received extracted data through channel and semantic decoding \cite{SCEI}. As an innovative approach, semantic communications have shown significant potential to alleviate spectrum shortage by reducing traffic loads over the network, garnering growing attention from both industry and academia \cite{Duan2023Multimedia,Qin2023Generalized}.

One important distinction from traditional communications is that semantic communication is a knowledge-based approach \cite{SC,Zhou2024Cognitive,xu2023knowledge,Zhang2023Deep,Hu2023Robust}.
The semantic transceivers need to construct their own knowledge bases (KBs) at initial stage by self-learning, like human brains, which form the foundation of semantic communications. All the necessary information that facilitates the communications at the semantic level is included in the KB. The transceivers may choose to use different classes of knowledge from their KBs according to different target tasks, scenarios and receivers.
Existing literature on KB construction can be categorized into three main categories: knowledge graph \cite{Zhou2024Cognitive,Wang2017Knowledge,xu2023knowledge}, training datasets \cite{Zhang2023Deep}, and semantic codebook \cite{Hu2023Robust}.
Transceiver designs in \cite{Zhou2024Cognitive,xu2023knowledge,Zhang2023Deep,Hu2023Robust} are based on the assumption that the accurate knowledge is shared between the KBs of transmitters and receivers, without considering the possible KB mismatch between the transmitter and receiver involved.

In semantic communications, the KBs have to be shared among transceivers in real time to ensure that the processes of semantic extraction and restoration can be exactly matched to obtain the meaningful information \cite{Liang2024Deep,Kadam2023ICC}. Since effective semantic communications depend on the perfect alignment of KBs between transmitters and receivers, mismatched knowledge would prevent extracted semantic information from accurate recovery and reconstruction, thus degrading the overall semantic communication performance. Hence, how to design effective semantic transmission mechanisms in the case of KB mismatch between the transmitter and receiver is an important and challenging problem to solve.
A hybrid semantic and bit communication network is considered in \cite{MS}, where each user can only select one communication mode. In semantic mode, a queuing model was developed, where both knowledge-matching and knowledge-mismatching packets are considered to derive the packet loss probability. In \cite{Sun2024Semantic}, a KB-enabled multi-level feature transmission framework was proposed, where different levels of features can be transmitted based on the KB matching conditions for remote zero-shot object recognition. To avoid KB mismatch, authors in \cite{KBC} proposed a KB construction scheme between vehicle-to-vehicle (V2V) pairs in vehicular networks, where a roadside unit (RSU) owns the entire KB directory and the KBs for all V2V pairs are constructed by downloading the required knowledge from the RSU to realize knowledge matching. However, the KBs between the transmitter and receiver could be different due to network dynamics.
These work \cite{KBC,MS,Sun2024Semantic} considers different transmission schemes to tackle the challenge of KB mismatch between transmitters and receivers, rather than considering the knowledge updating and sharing between transmitters and receivers.
Knowledge updating and sharing are crucial in semantic communications to maintain the perfect alignment of KBs between transmitters and receivers \cite{GAI,Lin2023Efficient,Chai2021Hierarchical,Du2023AI}. Updating KBs is critical to enable adaptation to new tasks and user demands, and further enhance personalized service and dedicated user experience.
\cite{Lin2023Efficient,Chai2021Hierarchical} considered knowledge sharing among edge servers utilizing collective learning and cooperative knowledge construction, which are usually implemented by federated learning. \cite{Lin2023Efficient,Du2023AI} employed periodic audits to identify and delete outdated or incorrect data in KBs, while concurrently integrating new knowledge. Nevertheless, these work only discusses the general knowledge updating and sharing schemes rather than in specific scenarios of semantic communications.

Unlike traditional data-oriented communications attempting to transmit every single bit, task-oriented semantic communications aim to only deliver task-related semantic information \cite{task1,task2,task3}, which have great potential to improve the efficiency of communication and reduce the essential KBs.
One crucial factor in task-oriented semantic communications is semantic extraction ratio, which should be designed based on specific target task requirements as it can decide the semantic transmission accuracy, wireless transmit data size and task computing loads, thus affecting the semantic communication effectiveness and task completion time.
The recent work \cite{Xu2024Task,Zheng2023Computing,Liu2024Adaptable,Zhang2023DRL} has investigated the resource allocation problems considering semantic extraction ratio in task-oriented semantic communications. A joint computing offloading and semantic compression scheme was investigated in mobile edge computing systems \cite{Zheng2023Computing} with task delay. The authors in \cite{Liu2024Adaptable} optimized semantic extraction ratio, power and bandwidth allocations, and user selection to maximize task success probability. A dynamic resource allocation scheme using deep reinforcement learning was proposed in \cite{Zhang2023DRL} to jointly optimize the semantic compression ratio, transmit power, and bandwidth of each user.
These work \cite{Xu2024Task,Zheng2023Computing,Liu2024Adaptable,Zhang2023DRL} assumes that the mobile users have the same shared knowledge with the edge server. However, in practice, the KBs at the mobile users and edge server could be mismatched as a result of network dynamics, which would eventually degrade the task performance.
Moreover, the knowledge updating and sharing between transmitters and receivers will also compete for the limited network resources \cite{cost}.
Thus, intergrading the designs of knowledge updating and sharing and semantic extraction ratio is significantly important in a resource-constrained network to overcome the case of KB mismatch while satisfying task requirements, which is not investigated in existing literature.

The aforementioned work considers one cell with multiple users and did not study the case of multi-cell networks.
Considering the channel conditions, computing capabilities, and constructed KBs at different small base stations (SBSs) could be distinct, the mobile devices (MDs) covered by multiple SBSs could access the best SBS. However, the wireless and computing resources at each SBS are limited, making it challenging to accommodate all the MDs. Due to the heterogeneity of transceivers and target task delay constraints of MDs, knowledge updating and sharing between both ends cannot always be guaranteed. Thus, how to design effective and efficient semantic transmission mechanisms in the case of KB mismatch between heterogeneous transceivers with limited resources, satisfying diverse service requirements of MDs is a significant problem to solve.

Motivated by the above, in this paper, we design a novel proactive knowledge sharing-enabled task-oriented hybrid semantic and bit transmission mechanism in a heterogenous multi-cell system with limited resources in order to tackle the KB mismatch problem, by jointly considering the knowledge updating and sharing, semantic extraction ratio, and SBS association. A generalized effective semantic transmission rate of hybrid semantic and bit communications is derived and maximized for all MDs under the semantic accuracy requirements and delay tolerances of MD target tasks.
The formulated problem is a mixed integer nonlinear programming (MINLP) problem, which is hard to be solved by traditional optimization methods. The problem is decomposed into multiple subproblems equivalently. One optimum algorithm and another efficient algorithm are further developed.

The main contributions of the paper are summarized below:
\vspace{-4.3mm}
\begin{itemize}
  \item We propose a novel knowledge sharing-enabled task-oriented hybrid semantic and bit transmission mechanism to tackle the KB mismatch problem, where a MD can proactively share and upload the task-related mismatched knowledge to the associated SBS. Considering the heterogeneous transceivers in multi-cell networks, delay tolerances, and channel conditions, MDs may not be able to upload all the mismatched knowledge. The bit communications are adopted as an aid to transmit the rest data related to unshared mismatched knowledge to guarantee the effective execution of target tasks.
  \item We derive a generalized effective semantic transmission rate of hybrid semantic and bit communications, which is maximized for all MDs by jointly optimizing the knowledge sharing portion, extraction ratio, and SBS association, while satisfying the semantic accuracy requirements and delay tolerances of MD target tasks.
  \item The formulated problem is an MINLP problem, which is decomposed into multiple joint knowledge updating and extraction ratio (KUER) subproblems and an SBS association subproblem. An optimum algorithm to the KUER subproblems is proposed and an efficient algorithm is further developed based on hierarchical class partitioning and monotonic optimization. The results of these subproblems are then fed into the SBS association subproblem, which is converted into an optimum matching problem and solved by modified Kuhn-Munkres (K-M) algorithm optimally.
  \item A variety of results are presented that characterize the tradeoffs between task delay tolerance, communication and computation resources, and knowledge data size in terms of total semantic rate. Our results show the proposed efficient solution can achieve close-to-optimum performance and outperform the comparisons for a wide range of system parameters.
\end{itemize}

The remainder of the paper is organized as follows. The system model and problem formulation is described in Section \ref{sec:systemmodel}. Following this in Section \ref{sec:solution}, an optimum algorithm is first proposed and then an efficient solution is developed to solve the optimization problem efficiently. In Section \ref{sec:simulation}, simulation results that demonstrate the quality of proposed solutions are given. Finally, we present the conclusions of the work in Section \ref{sec:conclusions}.

{\it Notations}: We define ${\cal A} \subseteq {\cal B}$ as ${\cal A}$ is a subset of ${\cal B}$, i.e., ${\cal A}$ and ${\cal B}$ are sets and every element of ${\cal A}$ is also an element of ${\cal B}$. ${\cal A}$ is a strict subset of ${\cal B}$, denoted by $ {\cal A} \subsetneq B$, i.e., ${\cal A}$ is a subset of ${\cal B}$ but ${\cal A}$ is not equal to ${\cal B}$ (there exists at least one element of ${\cal B}$ which is not an element of ${\cal A}$). ${\cal A} \setminus {\cal B}$ is defined to be the set which is comprised of elements in ${\cal A}$ that are not in ${\cal B}$. The union of ${\cal A}$ and ${\cal B}$ is denoted by ${\cal A} \cup {\cal B}$ and is the set of all elements of ${\cal A}$ or ${\cal B}$ or both. The empty set is denoted by $\emptyset$ that has no elements. We set $\sum_{a = A+1}^{A}{b_{a}}=0$. $\mathbb{R}_{+}$ denotes the set of non-negative real numbers.


\vspace{-3mm}
\section{System Model and Problem Formulation}
\label{sec:systemmodel}

\begin{figure}[t]
  \centering
  \includegraphics[height=53mm,width=68mm]{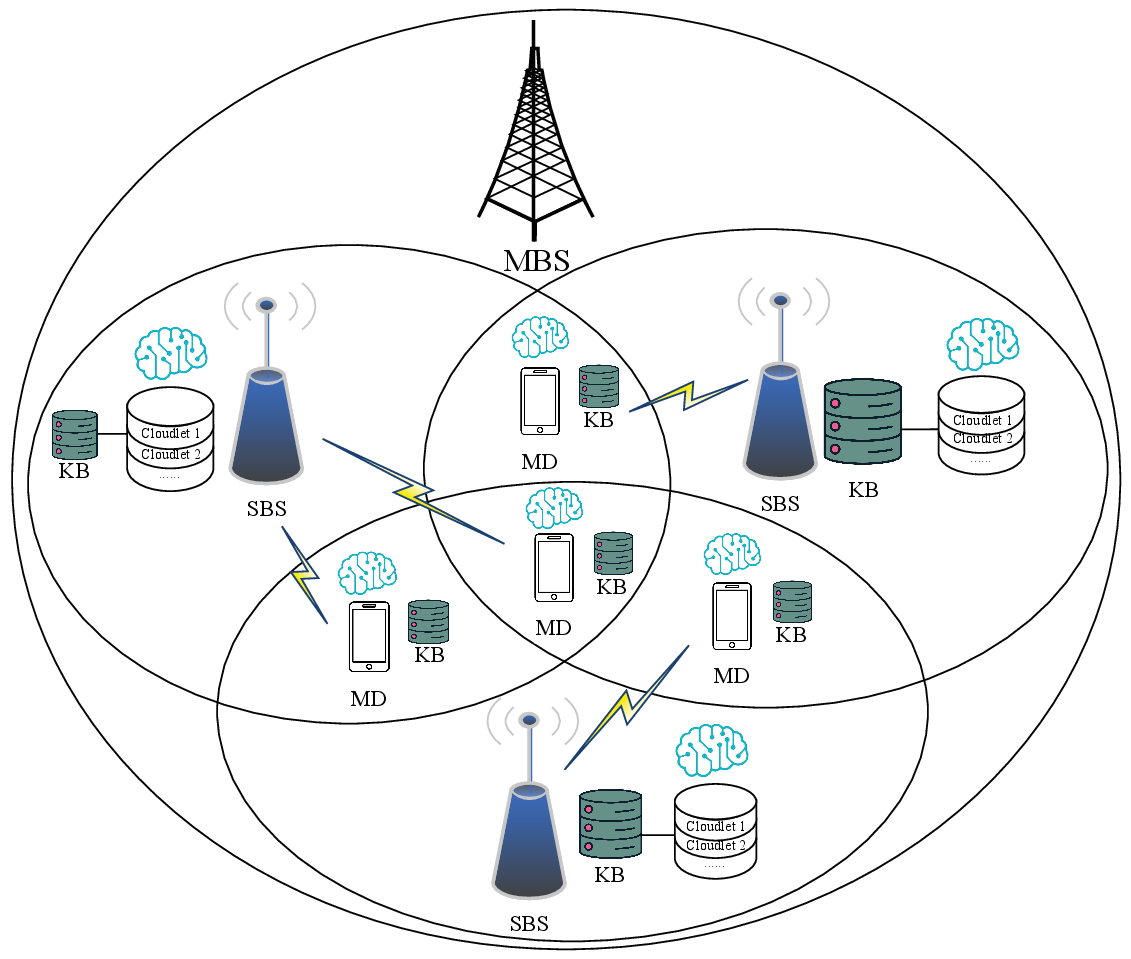}
  \caption{Network architecture for multi-cell KB-aware task-oriented sementic/bit communications.}
  \label{fig:1}
\end{figure}

As shown in Fig.~\ref{fig:1}, we consider a multi-cell multi-user network, consisting of $N$ SBSs coexisting within the coverage of a macro base station (MBS) and $M$ MDs within the coverage of multiple SBSs. The MBS serves as a central coordinator that can communicate with the SBSs through reliable backhaul links. SBS is indexed by $n \in {\cal N}=\{1,2,\ldots,N\} $. MD is indexed by $m \in {\cal M}=\{1,2,\ldots,M\}$. Both SBSs and MDs can perform semantic communications based on the shared knowledge to reduce the transmission load. In this case, each MD is equipped with a pre-trained semantic encoder so that it can transmit the extracted task-related semantic information of requested data to execute the target task at associated SBS. A KB is formed at each MD to store common and private knowledge, which could be some side information of the target task and codebook involved in the semantic encoder and decoder. On the other hand, each SBS is equipped with a pre-trained semantic decoder to recovery the received semantic information, and several cloudlet servers to execute the target tasks of MDs. A KB is also constructed at each SBS and needs to be updated based on the task-related knowledge from MDs.
We consider the MDs in the coverage of multiple SBSs, each of which needs to associate with one of the SBSs to transmit the requested data to the cloudlet server of the SBS to complete its target task. Define $x_{m,n}\in \{0,1\}$ as the binary variable representing if MD $m$ is associated with SBS $n$. If $x_{m,n}=1$, MD $m$ is associated with SBS $n$; otherwise, $x_{m,n}=0$.

\vspace{-3mm}
\subsection{Knowledge sharing-enabled hybrid transmission mechanism}
\label{sec:KS}

\begin{figure}[t]
  \centering
  \includegraphics[width=0.35\textwidth]{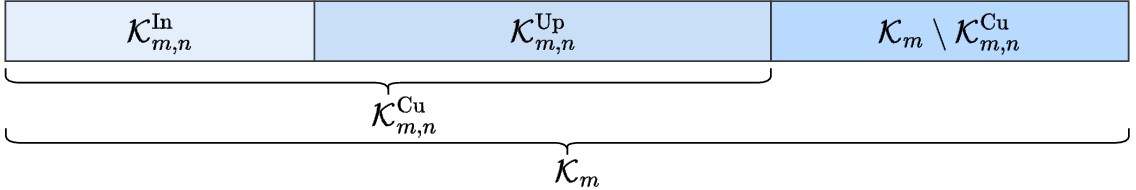}
  \caption{An illustration for knowledge class sets relationships when initial stored knowledge class set at SBS $n$ requested by MD $m$ ${\cal K}_{m,n}^{\rm In} \subsetneq {\cal K}_m$.}
  \label{fig:10}
  \vspace{-5mm}
\end{figure}

The effectiveness of semantic communications depend on the knowledge alignment between transmitters and receivers. In order to tackle the KB mismatch problem, we propose a novel proactive knowledge sharing-enabled semantic and bit transmission mechanism.
We define ${\cal K}_m=\{1,2,\ldots,K_m\} $ as the class set of task-related knowledge in $K_m$ classes at MD $m$, which is indexed by $k \in {\cal K}_m$. As SBSs have their own KBs, some classes of MD task-related knowledge may be already stored at the KBs of SBSs, e.g., some common knowledge classes. The initial set of stored task-related knowledge classes of MD $m$ at SBS $n$ is denoted by ${\cal K}_{m,n}^{\rm In} = \{1,2,\ldots,{K}_{m,n}^{\rm In}\}$.
If ${\cal K}_{m,n}^{\rm In} = {\cal K}_m$, MD $m$ can start semantic communications immediately with SBS $n$.
Otherwise, if ${\cal K}_{m,n}^{\rm In} \subsetneq {\cal K}_m$, the task-related knowledge in classes ${\cal K}_m \setminus {\cal K}_{m,n}^{\rm In}$ at MD $m$ needs to be shared to the associated SBS $n$ to perform effective semantic communications.
Define ${\cal K}_{m,n}^{\rm Up} = \{1,2,\ldots,{K}_{m,n}^{\rm Up}\} \subseteq {\cal K}_m \setminus {\cal K}_{m,n}^{\rm In}$ as the set of knowledge classes to be updated and shared from MD $m$ to SBS $n$. Hence, the current set of matched task-related knowledge classes after updating KB at associated SBS $n$ is ${\cal K}_{m,n}^{\rm Cu} = \{1,2,\ldots,{K}_{m,n}^{\rm Cu}\}$, where ${\cal K}_{m,n}^{\rm Cu}={\cal K}_{m,n}^{\rm In} \cup {\cal K}_{m,n}^{\rm Up}$. An illustration for the knowledge class sets relationships is given in Fig.~\ref{fig:10}.
However, due to the target task delay tolerance and random channel conditions, all the mismatched task-related knowledge of MDs may not be able to be uploaded to the KBs of associated SBSs.
That is, in the case of ${\cal K}_{m,n}^{\rm Cu} \subsetneq {\cal K}_m$, the requested data related to knowledge class $k\in {\cal K}_{m,n}^{\rm Cu}$ can be transmitted in semantic communications, while the rest data related to knowledge class $k \in {\cal K}_m \setminus {\cal K}_{m,n}^{\rm Cu}=\{{K}_{m,n}^{\rm Cu}+1,\ldots, K_m\}$ has to be transmitted in traditional bit communications as an aid.
We define $d_{m,k}^{\rm T}$ and $d_{m,k}^{\rm K}$ as the amount of requested raw data of target task and related knowledge data at MD $m$ in knowledge class $k$, respectively. Let $I_{m,k}$ and $c_{m,k}$ be the amount of semantic information and computation load of the requested data related to knowledge class $k$ of MD $m$, correspondingly.

\begin{figure}[t]
  \centering
  \includegraphics[width=0.4\textwidth]{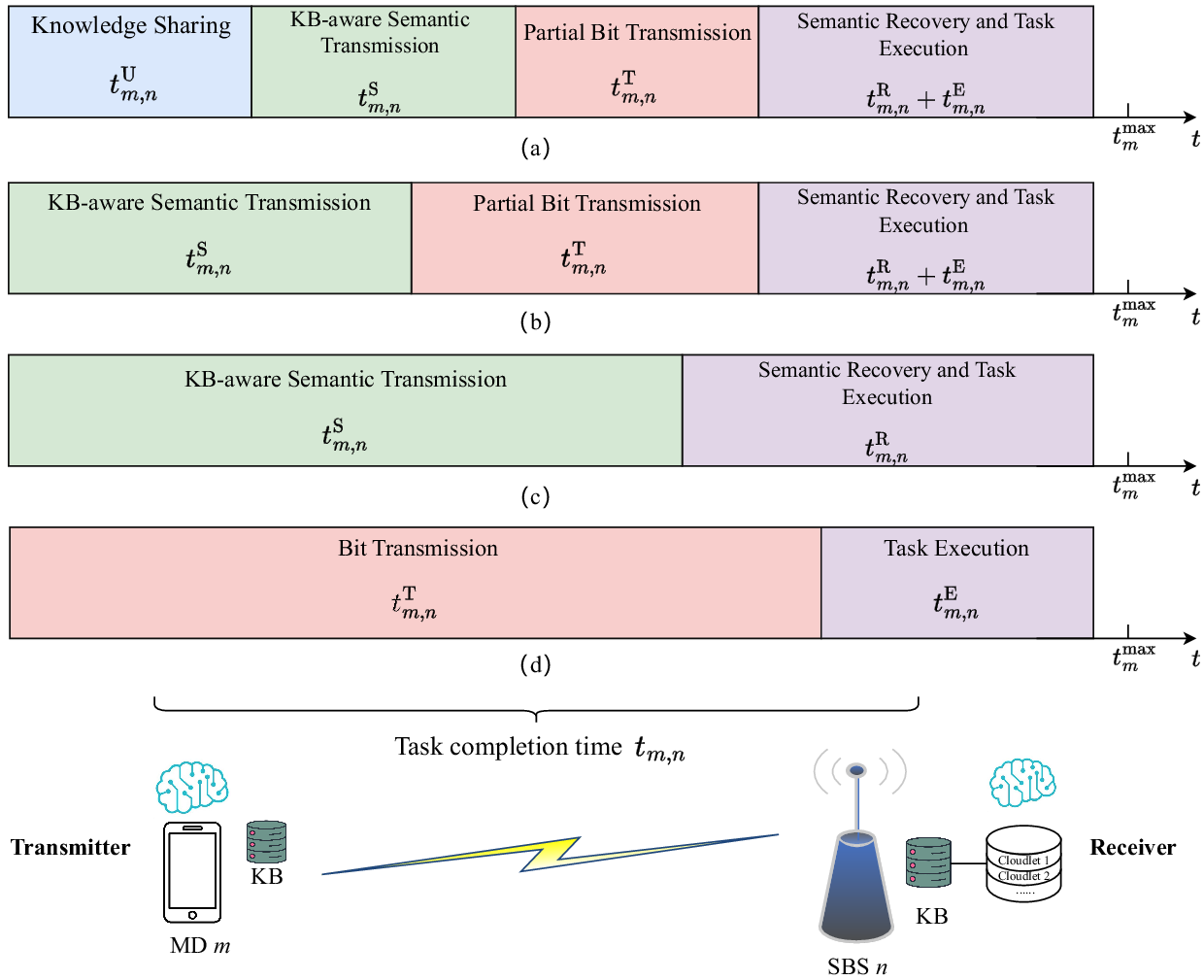}
  \caption{Procedure of proposed task-oriented semantic/bit communications. (a) semantic/bit communications with mismatched knowledge and knowledge sharing; (b) semantic/bit communications with mismatched knowledge without knowledge sharing; (c) semantic communications with well-matched knowledge; (d) bit communications without matched knowledge.}
  \label{fig:2}
\vspace{-5mm}
\end{figure}

Therefore, the transmission mechanism can be summarized into the following four cases, as shown in Fig.~\ref{fig:2}:
\begin{enumerate}
  \item When ${\cal K}_{m,n}^{\rm In} \subsetneq {\cal K}_m$ and ${\cal K}_{m,n}^{\rm Up} \neq \emptyset$, as shown in Fig.~\ref{fig:2}(a), it indicates that the mismatched task-related knowledge of MD $m$ is uploaded and shared to associated SBS $n$ to serve further semantic communications in between. Specially, the updated knowledge class set at associated SBS ${\cal K}_{m,n}^{\rm Cu} \subsetneq {\cal K}_m$, so that the requested data related to mismatched knowledge class $k\in {\cal K}_m \setminus {\cal K}_{m,n}^{\rm Cu} $ can only be transmitted in bit communications; if ${\cal K}_{m,n}^{\rm Cu} = {\cal K}_m$, the knowledge at SBS $n$ is well-matched with the task-related knowledge at MD $m$, so that all the requested data of target task from MD $m$ can be effectively transmitted in semantic communications.
  \item When ${\cal K}_{m,n}^{\rm In} \subsetneq {\cal K}_m$ and ${\cal K}_{m,n}^{\rm Up} = \emptyset$, ${\cal K}_{m,n}^{\rm Cu} ={\cal K}_{m,n}^{\rm In}$. As shown in Fig.~\ref{fig:2}(b), it reveals that the mismatched task-related knowledge of MD $m$ cannot be shared to associated SBS $n$ due to the large size of knowledge data. If so, it will degrade the overall performance. Thus, the MD starts semantic communications directly based on the initial matched knowledge, while the rest data related to the mismatched knowledge has to be transmitted in a traditional way.
  \item When ${\cal K}_{m,n}^{\rm In} = {\cal K}_m$, as shown in Fig.~\ref{fig:2}(c), it shows the task-related knowledge stored at SBS $n$ is well-matched with that at MD $m$, so that MD $m$ can start semantic communications immediately with SBS $n$. In this case, ${\cal K}_{m,n}^{\rm Up} = \emptyset$ and ${\cal K}_{m,n}^{\rm Cu} ={\cal K}_{m,n}^{\rm In} = {\cal K}_m$.
  \item When ${\cal K}_{m,n}^{\rm In} = {\cal K}_{m,n}^{\rm Up} = {\cal K}_{m,n}^{\rm Cu}= \emptyset$, as shown in Fig.~\ref{fig:2}(d), it indicates the KB at SBS $n$ does not store any task-related knowledge of MD $m$ and the mismatched knowledge of MD $m$ can not be uploaded and shared to SBS $n$. For example, if the size of knowledge data is too large, it would degrade the overall performance if transmitted, compared to direct bit communications. In this case, the MD prefers bit communications, and semantic decoding and recovery are no longer needed at the receiver.
\end{enumerate}

\vspace{-3mm}
\subsection{Task completion procedure}
\label{sec:UA}

In order to complete the target tasks, the requested data from MDs related to matched knowledge can be transmitted to associated SBSs in semantic communications. Meanwhile, bit communications as a complemental way can be adopted to transmit the data related to mismatched knowledge.
In case 1, ${\cal K}_{m,n}^{\rm In} \subsetneq {\cal K}_m$ and ${\cal K}_{m,n}^{\rm Up} \neq \emptyset$, ${\cal K}_{m,n}^{\rm Cu} ={\cal K}_{m,n}^{\rm In} \cup {\cal K}_{m,n}^{\rm Up}$.
For simplifying the calculation, we put set ${\cal K}_{m,n}^{\rm In}$ in the front of set ${\cal K}_{m,n}^{\rm Cu}$, followed by set ${\cal K}_{m,n}^{\rm Up}$, i.e., ${\cal K}_{m,n}^{\rm Cu} = \{1,2,\ldots,{K}_{m,n}^{\rm In},{K}_{m,n}^{\rm In}+1, \ldots, {K}_{m,n}^{\rm Cu}\}$.
Since the physical transmission still follows the Shannon theory, the task-related knowledge sharing transmission time $t_{m,n}^{\rm U}$ of classes in ${\cal K}_{m,n}^{\rm Up}$ from MD $m$ to SBS $n$ can be formulated as
\begin{equation}\label{KBT}
  t_{m,n}^{\rm U} = \frac{\sum\limits_{k = K_{m,n}^{\rm In}+1}^{K_{m,n}^{\rm Cu}}{d_{m,k}^{\rm K}}}{R_{m,n}},
\end{equation}
where $R_{m,n}$ is the data transmission rate from MD $m$ to SBS $n$, which is given by
\begin{equation}\label{Rmn}
  R_{m,n} = W{\log _2}( 1 + \frac{{p^{\rm{T}}_m}{g_{m,n}}}{\sigma^2}),
\end{equation}
where $W$ denotes the pre-allocated subchannel bandwidth to transmit the requested data, $p^{\rm{T}}_m$ and $g_{m,n}$ are the transmission power and the link gain from MD $m$ to SBS $n$, correspondingly, and $\sigma^2$ is the noise power at the SBS receiver input. Obviously, in other cases, $t_{m,n}^{\rm U}=0$. For example, in case 2, ${\cal K}_{m,n}^{\rm In} \subsetneq {\cal K}_m$ but ${\cal K}_{m,n}^{\rm Up} = \emptyset$, thus ${\cal K}_{m,n}^{\rm Cu} ={\cal K}_{m,n}^{\rm In}$ and $t_{m,n}^{\rm U}=0$.

The requested data of MDs related to matched knowledge classes can be semantically extracted and transmitted to associated SBSs for further task execution. Let $\xi_{m,n}\in [0,1]$ be the semantic extraction ratio of requested data of MD $m$ associated with SBS $n$. The smaller extraction ratio means higher compression rate. The semantic transmission time $t_{m,n}^{\rm S}$ from MD $m$ to SBS $n$ can be expressed as
\begin{equation}
  t_{m,n}^{\rm S} = \frac{\xi_{m,n} \sum\limits_{k = 1}^{K_{m,n}^{\rm Cu}} {d_{m,k}^{\rm T}}}{R_{m,n}}.
\end{equation}

If current knowledge class set ${\cal K}_{m,n}^{\rm Cu} \subsetneq {\cal K}_m$, the rest requested data related to mismatched knowledge classes can only be transmitted in bit communications. The bit transmission time $t_{m,n}^{\rm T}$ from MD $m$ to SBS $n$ can be formulated as
\begin{equation} \label{tT}
  t_{m,n}^{\rm T} = \frac{\sum\limits_{k = K_{m,n}^{\rm Cu}+1}^{K_{m}}{d_{m,k}^{\rm T}}}{R_{m,n}}.
\end{equation}
Otherwise, if ${\cal K}_{m,n}^{\rm Cu} = {\cal K}_m$, all the data can be transmitted in semantic communications, so that $t_{m,n}^{\rm T}=0$.

In case 3, ${\cal K}_{m,n}^{\rm Up} = \emptyset$, so that ${\cal K}_{m,n}^{\rm Cu} ={\cal K}_{m,n}^{\rm In} = {\cal K}_m$, $t_{m,n}^{\rm U}= t_{m,n}^{\rm T} =0 $ and the MD starts semantic transmissions immediately.
In case 4, ${\cal K}_{m,n}^{\rm In} = {\cal K}_{m,n}^{\rm Up} = {\cal K}_{m,n}^{\rm Cu}= \emptyset$, thus $t_{m,n}^{\rm U}=t_{m,n}^{\rm S}=0$ and $t_{m,n}^{\rm T} = \frac{\sum_{k = 1}^{K_{m}}{d_{m,k}^{\rm T}}}{R_{m,n}}$, which is consistent with \eqref{tT}. The MD only performs bit communications.
Note that since the semantic encoders and decoders are pre-trained models, the task-related knowledge is used in fine-tuning the models, which is fast in general. Hence, the time for model fine-tuning can be ignored compared to the wireless data transmission and task execution time.

When an SBS receives the data from an associated MD, it will assign one of the cloudlet servers to execute the MD target task. If the requested data is transmitted in semantic communications, additional computation load is required to process the semantic data rather than raw data. Since smaller extraction ratio introduces more computation loads for semantic decoding and reconstruction, we define $\omega_{m,n} \ge 1$ as the ratio of computation loads of semantic data to that of raw data from MD $m$ to SBS $n$. It increases monotonously with semantic extraction ratio $\xi_{m,n}$ getting small \cite{Cang2023Online}. $\omega_{m,n} = 1$ if raw data is transmitted, i.e., $\xi_{m,n} = 1$. Without loss of generality, we assume that
\begin{equation}\label{cl}
  \omega_{m,n}= \frac{1}{{\xi_{m,n}}^{\rho}},
\end{equation}
where $\rho > 0$ is a constant parameter varying with the semantic models relevant to different task types. Hence, the computing time when receiving semantic data to execute MD target task can be formulated as
\begin{equation}\label{CT1}
  t_{m,n}^{\rm R} = \frac{\omega_{m,n} \sum\limits_{k = 1}^{K_{m,n}^{\rm Cu}} {c_{m,k}}}{f_n^{\rm C}},
\end{equation}
where ${f_n^{\rm C}}$ is the computing speed of a cloudlet at SBS $n$ in number of CPU cycles per second. While the computing time when receiving raw data to execute the task can be given by
\begin{equation}\label{CT2}
  t_{m,n}^{\rm E} = \frac{\sum\limits_{k = K_{m,n}^{\rm Cu}+1}^{K_{m}} {c_{m,k}}}{f_n^{\rm C}}.
\end{equation}
Especially, when ${\cal K}_{m,n}^{\rm Cu} = {\cal K}_m$, the received data at SBS $n$ are all semantic data, so that $t_{m,n}^{\rm E}=0$. When ${\cal K}_{m,n}^{\rm In} = {\cal K}_{m,n}^{\rm Up} = {\cal K}_{m,n}^{\rm Cu}= \emptyset$, the received data at SBS $n$ are all raw data, so that $t_{m,n}^{\rm R}=0$.

As a result, the total task completion time of MD $m$ at a cloudlet of SBS $n$ can be obtained
\begin{equation}\label{TT}
  t_{m,n} = t_{m,n}^{\rm U} +t_{m,n}^{\rm S} + t_{m,n}^{\rm T} + t_{m,n}^{\rm R} + t_{m,n}^{\rm E}.
\end{equation}
Each MD $m$ requires that the target task must be completed in maximum delay tolerance $d_m^{\max}$. Hence, the following delay constraint must be satisfied for all $m$, i.e.,
\begin{equation}\label{Delay}
  \sum\limits_{n = 1}^N {x_{m,n} t_{m,n}} \le t_m^{\max}.
\end{equation}

\subsection{SBS associations in multi-cell networks}
\label{sec:UA}

In order to execute the target task, each MD can associate one and only one SBS to transmit the requested information by semantic and bit communications. Thus, the following constraint must hold,
\begin{equation}\label{UA}
  \sum\limits_{n = 1}^N x_{m,n} \le 1,
\end{equation}
for all $m \in {\cal M}$.
Rather than considering dynamic computation resource allocations through multitasking or virtual machine allocations that cost extra overhead \cite{Chen2020Joint}, we consider that each SBS has multiple cloudlet servers (or CPU cores), each of which has a fixed computing speed. When an SBS receives the requested data of target task from associated MD, it assigns one of the cloudlet servers to process the requested task. The total number of cloudlet servers at SBS $n$ is denoted by $S_n^{\max}$. Thus, the total number of MDs associated with SBS $n$ to execute their target tasks is limited, i.e.,
\begin{equation} \label{BA}
\sum\limits_{m = 1}^M x_{m,n} \le S_n^{\max}.
\end{equation}

\subsection{Semantic transmission accuracy and generalized effective semantic transmission rate}
\label{sec:SA}

Semantic extraction ratio $\xi_{m,n}$ is a critical factor affecting semantic transmission accuracy $\eps_{m,n}(\xi_{m,n})$ between MD $m$ and SBS $n$. However, a closed-form formula of $\eps_{m,n}(\xi_{m,n})$ is intractable to be derived due to the unexplainability of neural networks of semantic models. Fortunately, the relationship between the semantic accuracy and extraction ratio can be approximated by a function to reflect the statistics of the model evaluation. In general, the semantic accuracy $\eps_{m,n}(\xi_{m,n})$ can be characterized as a nonlinear function of semantic extraction ratio $\xi_{m,n}$ \cite{Xu2024Task,Wang2020Machine,Liu2024Adaptable}. Particularly, this nonlinear function should satisfy the following properties:
\begin{enumerate}
  \item As $\eps_{m,n}$ is a percentage, it satisfies $0\le \eps_{m,n}(\xi_{m,n}) \le 1$;
  \item The smaller the semantic extraction ratio is, the worse the semantic accuracy is. Since the semantic information of raw data is transmitted with less data, the reconstructed information is less accurate. Thus, the semantic accuracy is a monotonically increasing function of semantic extraction ratio.
\end{enumerate}
Based on these properties, $\eps_{m,n}(\xi_{m,n})$ can be estimated by nonlinear model $\eps'_{m,n}(\xi_{m,n}|\theta_1,\theta_2,\theta_3,\theta_4)$, which is given by
\begin{align}\label{accuracymodel}
  \eps_{m,n}(\xi_{m,n}) &\approx \eps'_{m,n}(\xi_{m,n}|\theta_1,\theta_2,\theta_3,\theta_4) \nonumber \\
  &=-\theta_1  e^{(\theta_2 (1-\xi_{m,n}))}+\theta_3 e^{(-\theta_4 (1-\xi_{m,n}))},
\end{align}
where $\theta_1,\theta_2,\theta_3,\theta_4 \ge 0$ are tuning parameters. With given $\{\xi_{m,n}^{(i)}, \eps_{m,n}(\xi_{m,n}^{(i)}) \}_{i=1}^Q$, where $Q$ is the number of points to be fitted, the optimal parameters $(\theta_1,\theta_2,\theta_3,\theta_4)$ can be found via the nonlinear least squares fitting:
\begin{align}\label{fitting}
  \{\theta_1,\theta_2,\theta_3,\theta_4\}&=\arg\min_{\theta_1,\theta_2,\theta_3,\theta_4} \nonumber \\
  \frac{1}{Q} &\sum\limits_{i=1}^{Q}|\eps_{m,n}(\xi_{m,n}^{(i)})-\eps'_{m,n}(\xi_{m,n}^{(i)}|\theta_1,\theta_2,\theta_3,\theta_4)|^2.
\end{align}
The Levenberg–Marquardt method \cite{gavin2019levenberg} can be used for nonlinear curve fitting. The values of tuning parameters $(\theta_1,\theta_2,\theta_3,\theta_4)$ vary with adopted semantic models relevant to different task types.

Each MD $m$ has a minimum semantic accuracy requirement $\eps_m^{\rm th}$ to guarantee semantic transmission performance, so that the following semantic accuracy constraint must hold for all $m$, i.e.,
\begin{equation}\label{accuracy}
  \sum\limits_{n = 1}^N x_{m,n} \eps_{m,n}(\xi_{m,n}) \ge \eps_m^{\rm th}.
\end{equation}

The effective semantic transmission rate between MD $m$ and SBS $n$ is denoted by $\gamma_{m,n}$. Unlike bit-stream data rate, the semantic unit (sut) as the basic unit of semantic information can be used to measure the amount of semantic information \cite{Yan2022Resource,Chen2024Semantic}. Thus, effective semantic transmission rate, as one of the crucial semantic-based performance metrics, is defined as the effectively transmitted semantic information per second in suts/s. In our proposed transmission mechanism, the requested data of target task of MD $m$ could be transmitted in two communication manners, i.e., semantic and bit communications.
The requested data related to knowledge class $k$ in current matched knowledge class set ${\cal K}_{m,n}^{\rm Cu}$ is transmitted in semantic communications; while the rest requested date of MD $m$ is transmitted to SBS $n$ in bit communications due to the mismatch of task-related knowledge.
Therefore, the generalized effective semantic transmission rate from MD $m$ to SBS $n$ can be formulated as
\begin{equation}\label{STR}
  \gamma_{m,n}= \frac{R_{m,n}(\sum\limits_{k = 1}^{K_{m,n}^{\rm Cu}} I_{m,k} ~\eps_{m,n}(\xi_{m,n})+\!\!\sum\limits_{k = K_{m,n}^{\rm Cu}+1}^{K_{m}}\!\!I_{m,k})}{\xi_{m,n}\sum\limits_{k = 1}^{K_{m,n}^{\rm Cu}} {d_{m,k}^{\rm T}} + \!\!\sum\limits_{k = K_{m,n}^{\rm Cu}+1}^{K_{m}} {d_{m,k}^{\rm T}}} .
\end{equation}
Note that when ${\cal K}_{m,n}^{\rm Cu}=\emptyset$, the requested data is transmitted from MD $m$ to SBS $n$ only in bit communications, so that the first term in the bracket in the numerator and the first term in the denominator are both zero. When ${\cal K}_{m,n}^{\rm Cu} = {\cal K}_m$, traditional bit communications are no longer needed so that the second term in the bracket in the numerator and the second term in the denominator are both zero.

When an MD is in the coverage of multiple SBSs, it can choose to associate with one SBS based on the KB, channel conditions, and computation capacities to perform knowledge sharing and corresponding semantic and bit communications to transmit the required data to execute its target task using one of the cloudlets.
Our goal is to maximize the total generalized effective semantic transmission rate of all MDs by jointly optimizing SBS association, knowledge sharing decision, and semantic extraction ratio, while considering semantic accuracy requirements and target task completion time constraints of MDs, and limited computation capacities of SBSs.
The optimization problem can be formulated as follows:
\begin{subequations} \label{Eq:maxtotal}
\begin{align}
&\max_{x_{m,n}, {\cal K}_{m,n}^{\rm Cu}, \xi_{m,n}}\sum\limits_{m = 1}^M\sum\limits_{n = 1}^N {x_{m,n}  \gamma_{m,n}({\cal K}_{m,n}^{\rm Cu}, \xi_{m,n}) }   \\
\text{s.t.} & \sum\limits_{n = 1}^N x_{m,n} (t_{m,n}^{\rm U}({\cal K}_{m,n}^{\rm Cu}) +t_{m,n}^{\rm S}({\cal K}_{m,n}^{\rm Cu}, \xi_{m,n}) + t_{m,n}^{\rm T}({\cal K}_{m,n}^{\rm Cu}) +  \nonumber \\
&~~~~~~~~~~~t_{m,n}^{\rm R}({\cal K}_{m,n}^{\rm Cu}, \xi_{m,n}) + t_{m,n}^{\rm E}({\cal K}_{m,n}^{\rm Cu})) \le t_m^{\max}, \forall m    \label{Eq:C1} \\
& \eqref{UA}, \eqref{BA}, \eqref{accuracy},   \label{Eq:C2} \\
& x_{m,n} \in \{0,1\}, \forall m,n \label{Eq:C3}\\
& \xi_{m,n} \in [0,1], \forall m,n \label{Eq:C4}\\
&{\cal K}_{m,n}^{\rm Cu} \subseteq {\cal K}_{m}, \forall m,n.   \label{Eq:C5}
\end{align}
\end{subequations}
Since SBS associations $x_{m,n}$'s and knowledge sharing decisions ${\cal K}_{m,n}^{\rm Cu}$'s are integer variables and the optimization variables are mutual coupled, problem \eqref{Eq:maxtotal} is a non-convex nonlinear problem and belongs to the family of MINLP problem. In general, the MINLP problems are NP-hard and cannot be solved efficiently using traditional optimization methods. Next, we will decompose it into separate subproblems equivalently and solve it efficiently using proposed efficient algorithm.

\vspace{-3mm}
\section{Proposed Solutions to the Problem}
\label{sec:solution}

Since problem \eqref{Eq:maxtotal} is an MINLP problem, it is difficult to find the optimal solution conveniently using traditional optimization tools. In this section, problem \eqref{Eq:maxtotal} is decomposed into multiple joint KUER subproblems, each of which is for one SBS and one of the potential MDs that may be associated to the SBS, and an SBS association subproblem. Each joint KUER subproblem can be classified into two cases based on the relationship of knowledge class sets ${\cal K}_{m}$ and ${\cal K}_{m,n}^{\rm In}$. If ${\cal K}_{m,n}^{\rm In}={\cal K}_{m}$, MD $m$ can transmit the requested data in semantic communications directly so that there is no need to optimize knowledge sharing decisions, and therefore, the subproblem is reduced into a non-convex nonlinear problem with variables $\xi_{m,n}$'s. If ${\cal K}_{m,n}^{\rm In} \subsetneq {\cal K}_{m}$, joint KUER subproblems are still MINLP problems. We propose an optimum algorithm to the subproblems used for small-scale scenarios. Due to the exponential complexity caused from the combinations of mismatched knowledge classes, a low-complexity algorithm is proposed to obtain near-optimal solution efficiently for large-scale scenarios.
The results of these subproblems are then fed into the SBS association subproblem, which becomes a linear binary problem, and can be converted into an optimum matching problem and solved by a modified K-M algorithm \cite{KM1} optimally.

\vspace{-3mm}
\subsection{Joint KUER subproblems}
\label{subsec:subproblem1}

Given $x_{m,n}=1$, i.e., assume MD $m$ is associated with SBS $n$, jointly solving ${\cal K}_{m,n}^{\rm Cu}$ and $\xi_{m,n}$ is an independent subproblem between MD $m$ and SBS $n$, which can be separated from optimization problem \eqref{Eq:maxtotal} and given as follows:
\begin{subequations} \label{Eq:maxtotalmn}
\begin{align}
&\!\!\!\!\!\max_{{\cal K}_{m,n}^{\rm Cu}, \xi_{m,n}}\!\!\! \frac{R_{m,n}(\sum\limits_{k = 1}^{K_{m,n}^{\rm Cu}} I_{m,k} ~\eps_{m,n}(\xi_{m,n})+\!\!\sum\limits_{k = K_{m,n}^{\rm Cu}+1}^{K_{m}}\!\!I_{m,k})}{\xi_{m,n}\sum\limits_{k = 1}^{K_{m,n}^{\rm Cu}} {d_{m,k}^{\rm T}} + \!\!\sum\limits_{k = K_{m,n}^{\rm Cu}+1}^{K_{m}} {d_{m,k}^{\rm T}}} \\
&\text{s.t.} ~~t_{m,n}^{\rm U}({\cal K}_{m,n}^{\rm Cu}) +t_{m,n}^{\rm S}({\cal K}_{m,n}^{\rm Cu}, \xi_{m,n}) + t_{m,n}^{\rm T}({\cal K}_{m,n}^{\rm Cu}) +  \nonumber \\
&~~~~~t_{m,n}^{\rm R}({\cal K}_{m,n}^{\rm Cu}, \xi_{m,n}) + t_{m,n}^{\rm E}({\cal K}_{m,n}^{\rm Cu}) \le t_m^{\max}, \label{Eq:delay} \\
&~~~~~\eps_{m,n}(\xi_{m,n}) \ge \eps_m^{\rm th},   \label{Eq:xi} \\
%
%
%
%
&~~~~~\eqref{Eq:C4}, \eqref{Eq:C5}.
\end{align}
\end{subequations}
Especially, when ${\cal K}_{m,n}^{\rm In}={\cal K}_{m}$, there is no knowledge updating so that the joint subproblem is reduced into an optimization problem with only variables $\xi_{m,n}$'s. Besides, there are no bit communications as a complimentary transmission manner. That is,
\begin{subequations} \label{Eq:maxtotalmnE}
\begin{align}
\max_{\xi_{m,n}} & \frac{R_{m,n}\sum\limits_{k = 1}^{K_{m}} I_{m,k} ~\eps_{m,n}(\xi_{m,n})}{\xi_{m,n}\sum\limits_{k = 1}^{K_{m}} {d_{m,k}^{\rm T}} }      \\
\text{s.t.} &~t_{m,n}^{\rm S}(\xi_{m,n}) + t_{m,n}^{\rm R}(\xi_{m,n}) \le t_m^{\max}, \\
&~\eps_{m,n}(\xi_{m,n}) \ge \eps_m^{\rm th},   \\
%
%
%
%
&~\xi_{m,n} \in [0,1].
%
\end{align}
\end{subequations}
Problem \eqref{Eq:maxtotalmnE} is a nonlinear non-convex optimization problem due to the intractability of the closed-form formula of $\eps_{m,n}(\xi_{m,n})$. Thus, there is no convenient traditional optimization methods to solve it efficiently.
Thanks to the specific characteristics of $\eps_{m,n}(\xi_{m,n})$, we can develop an optimum algorithm to solve it in the following subsection.

For the case of ${\cal K}_{m,n}^{\rm In} \subsetneq {\cal K}_{m}$, optimization problem \eqref{Eq:maxtotalmn} is still an MINLP problem with variables ${\cal K}_{m,n}^{\rm Cu}$'s and $\xi_{m,n}$'s. Next, we will first propose an optimum algorithm to solve joint KUER subproblem. Due to the high computation complexity, we then design a low-complexity algorithm to obtain near-optimum solution efficiently.

\subsubsection{Optimum algorithm}
\label{subsubsec:optsolution}

Joint KUER subproblem \eqref{Eq:maxtotalmn} is challenging to be solved directly using conventional optimization methods for MINLP problems, such as integer relaxation. That is because the discrete variable, current matched knowledge class set ${\cal K}_{m,n}^{\rm Cu}$, shows in the limitations of summations in the formulations.
Since ${\cal K}_{m,n}^{\rm Cu}={\cal K}_{m,n}^{\rm In} \cup {\cal K}_{m,n}^{\rm Up}$ and ${\cal K}_{m,n}^{\rm In}$ is known, we need to optimize the knowledge sharing class set ${\cal K}_{m,n}^{\rm Up} \subseteq {\cal K}_m \setminus {\cal K}_{m,n}^{\rm In}$.
Thus, we can enumerate all the possible combinations of ${\cal K}_{m,n}^{\rm Up}$.
An optimum algorithm is proposed to solve the joint subproblem: for each given set ${\cal K}_{m,n}^{\rm Up}$, i.e., ${\cal K}_{m,n}^{\rm Cu}$ is given, subproblem \eqref{Eq:maxtotalmn} is reduced into a nonlinear non-convex optimization problem with only variables $\xi_{m,n}$'s. It will be demonstrated that can be transformed into a monotonic optimization problem and solved efficiently by Polyblock outer approximation algorithm optimally in Section \ref{subsec:effsolution}; After the reduced subproblem is solved, we keep the solution and move to the next possibility of ${\cal K}_{m,n}^{\rm Up}$ and do the above process again. It terminates when all the possibilities of ${\cal K}_{m,n}^{\rm Up}$ are enumerated and then outputs the maximum semantic transmission rate with optimal solution ${\cal K}_{m,n}^{\rm Cu*}, \xi_{m,n}^*$ to subproblem \eqref{Eq:maxtotalmn}.

\vspace{-3mm}
\subsection{Efficient algorithm to joint KUER subproblems}
\label{subsec:effsolution}

In joint KUER subproblem, our goal is to find the optimal matched knowledge class set and semantic extraction ratio, in order to maximize the semantic transmission rate between MD $m$ and SBS $n$. Specifically, the problem design is to determine to partition the total task-related knowledge class set ${\cal K}_{m}$ into sets ${\cal K}_{m,n}^{\rm Cu}$ and ${\cal K}_m \setminus {\cal K}_{m,n}^{\rm Cu}$, so that the mismatched knowledge classes in ${\cal K}_{m,n}^{\rm Up}={\cal K}_{m,n}^{\rm Cu}\setminus {\cal K}_{m,n}^{\rm In}$ are shared from MD $m$ to SBS $n$. The requested data of MD $m$ related to knowledge class $k \in {\cal K}_{m,n}^{\rm Cu}$ can be transmitted to SBS $n$ in semantic communications, while other requested data is automatically transmitted in bit communications. Semantic extraction ratio for requested data with knowledge classes in ${\cal K}_{m,n}^{\rm Cu}$ is further optimized.

Optimizing ${\cal K}_{m,n}^{\rm Cu}$ is equivalent to optimizing ${\cal K}_{m,n}^{\rm Up} \subseteq {\cal K}_m \setminus {\cal K}_{m,n}^{\rm In}$. Since the optimum algorithm needs to go through all $2^{(K_m-K_{m,n}^{\rm In})}+1$ subsets of ${\cal K}_m \setminus {\cal K}_{m,n}^{\rm In}$ in order to find the best ${\cal K}_{m,n}^{\rm Up}$, the computation complexity is exponential so that we aim to develop a low-complexity class partitioning algorithm to obtain a good solution efficiently. The main idea is to sort the $K_m-K_{m,n}^{\rm In}$ mismatched knowledge classes in ${\cal K}_m \setminus {\cal K}_{m,n}^{\rm In}$ into an ordered list $\beta_1, \beta_2, \ldots, \beta_{K_m-K_{m,n}^{\rm In}}$ according to a certain criterion, and then partition this list into subsets ${\cal K}_{m,n}^{\rm Up}=\{\beta_1,\beta_2, \ldots,\beta_{{K}_{m,n}^{\rm Up}}\}$ and $( {\cal K}_m \setminus {\cal K}_{m,n}^{\rm In}) \setminus {\cal K}_{m,n}^{\rm Up} = \{\beta_{{K}_{m,n}^{\rm Up}+1}, \beta_{{K}_{m,n}^{\rm Up}+2}, \ldots, \beta_{K_m-K_{m,n}^{\rm In}}\}$. The sorting criteria and method to determine ${K}_{m,n}^{\rm Up}$ are introduced in the following.

\begin{algorithm}[t]
\caption{Two-tier class partitioning} \label{algo:1}
\small{
\begin{algorithmic}[1]
\State Sort knowledge classes in ${\cal K}_m \setminus {\cal K}_{m,n}^{\rm In}$ in decreasing order of $\Phi_{m,k}^{\rm B}$; let $\beta_1, \beta_2, \ldots, \beta_{K_m-K_{m,n}^{\rm In}}$ be the resulted ordering  \label{line:start}
\State Sort knowledge classes with each same $\Phi_{m,k}^{\rm B}$ values in increasing order of $\Phi_{m,n,k}^{\rm D}$; let $\beta'_1, \beta'_2, \ldots, \beta'_{K_m-K_{m,n}^{\rm In}}$ be the final ordered list
\State $\gamma^{\max}=0$     \label{line:rstart}
\For {${K}_{m,n}^{\rm Up}=0$ to $ K_m-K_{m,n}^{\rm In}$}
	\State ${\cal K}_{m,n}^{\rm Up}=\{\beta'_1,\beta'_2, \ldots, \beta'_{{K}_{m,n}^{\rm Up}}\}$
	\State Solve reduced subproblem with $\xi_{m,n}$ and obtain maximum $\gamma_{m,n}$ \label{line:maxsub2}
	\If {$\gamma_{m,n} > \gamma^{\max}$}
		\State $\gamma^{\max} = \gamma_{m,n}$; ${K}_{m,n}^{\rm Up*}= {K}_{m,n}^{\rm Up}$; $\xi_{m,n}^*=\xi_{m,n}$
	\EndIf
\EndFor    \label{line:end}
\State\Return {${\cal K}_{m,n}^{\rm Up*}=\{\beta'_1,\beta'_2, \ldots, \beta'_{{K}_{m,n}^{\rm Up*}}\}$, $\xi_{m,n}^*$}   \label{line:return}
\end{algorithmic}
}
\end{algorithm}

\subsubsection{Class partitioning}
\label{subsubsec:CSP}

The proposed method to determine the knowledge class set to be shared ${K}_{m,n}^{\rm Up}$ from MD $m$ to SBS $n$ is implemented based on two-tier ordering criterions. Considering the benefits from semantic communications, MDs in our system intent to transmit the requested data of target task to the cloudlets of SBSs in semantic communications. In order to implement effective semantic communications, the MD needs to share the mismatched knowledge to the KB of associated SBS. Considering the communication overhead, the intuition behind the first tier sorting criterion is to share the mismatched knowledge class with smaller amount of knowledge data, but the amount of requested raw data related to this knowledge class is large. It indicates that a large amount of requested raw data can be transmitted in semantic communications while with small communication overhead for knowledge alignment. Thus, the first-tier sorting criterion is the ratio between the amount of requested raw data over the amount of knowledge data related to the same knowledge class, which can be defined as
\begin{equation}\label{firsttier}
  \Phi_{m,k}^{\rm B} = \frac{d_{m,k}^{\rm T}}{d_{m,k}^{\rm K}}.
\end{equation}
The knowledge classes in ${\cal K}_m \setminus {\cal K}_{m,n}^{\rm In}$ are ordered in decreasing values of $\Phi_{m,k}^{\rm B}$.
After this first-tier ordering, we adopt a second-tier sorting criterion for the knowledge classes with same $\Phi_{m,k}^{\rm B}$ values. These knowledge classes are further ordered according to a second-tier sorting criterion, which takes the task completion time into account and defined as follows:
\begin{align}\label{secondtier}
  \Phi_{m,n,k}^{\rm D} =& \frac{t_{m,n,k}^{\rm S}}{t_{m,n,k}^{\rm B}} \nonumber \\
  =&\frac{(d_{m,k}^{\rm K}+\xi_{m,n}^{\rm th} {d_{m,k}^{\rm T}})/ R_{m,n} + \frac{1}{{\xi_{m,n}^{\rm th}}^{\rho}}c_{m,k} / f_n^{\rm C}}{{d_{m,k}^{\rm T}}/ R_{m,n} +c_{m,k} / f_n^{\rm C}},
\end{align}
where $\Phi_{m,n,k}^{\rm D}$ is a ratio between the estimated completion time $t_{m,n,k}^{\rm S}$ in semantic communications over the estimated completion time $t_{m,n,k}^{\rm B}$ in bit communications for partial task completion related to knowledge class $k$ from MD $m$ to SBS $n$, $t_{m,n,k}^{\rm S}$ includes the knowledge uploading time, semantic transmission time, and semantic recovery and execution time, $t_{m,n,k}^{\rm B}$ includes the bit transmission time and execution time. In this case, we prefer to share the knowledge class with relatively smaller task completion time in semantic communications compared to that occurring in bit communications. Note that since semantic extraction ratio has not been optimized yet, we use the minimum extraction ratio obtained from semantic accuracy constraint as $\xi_{m,n}^{\rm th}$ to measure the estimated completion time in semantic communications. Thus, the knowledge classes with same $\Phi_{m,k}^{\rm B}$ values are further sorted in increasing values of $\Phi_{m,n,k}^{\rm D}$. This two-tier partitioning method is given in Algorithm~\ref{algo:1}.
The algorithm employs a linear search process \cite{Partitioning} for best ${K}_{m,n}^{\rm Up}$, starting from $0$ and going up to $K_m-K_{m,n}^{\rm In}$.
Note that when ${K}_{m,n}^{\rm Up}=0$, no knowledge classes are allowed to shared from MD $m$ to SBS $n$; and when ${K}_{m,n}^{\rm Up}=K_m-K_{m,n}^{\rm In}$, all the mismatched knowledge classes are allowed to be shared to SBS $n$. For a given ${K}_{m,n}^{\rm Up}$, the current matched knowledge class set after sharing ${\cal K}_{m,n}^{\rm Cu}={\cal K}_{m,n}^{\rm In} \cup {\cal K}_{m,n}^{\rm Up}$ is known, so that the joint subproblem is reduced into a nonlinear non-convex problem with only $\xi_{m,n}$, which can be converted into a monotonic optimization problem. The algorithm returns ${\cal K}_{m,n}^{\rm Up *}$ that achieves the maximum semantic transmission rate among these solutions.

\subsubsection{Monotonic optimization}
\label{subsubsec:MO}

With given ${\cal K}_{m,n}^{\rm Cu}$, the joint KUER subproblem is reduced into a problem with only variable $\xi_{m,n}$, i.e., subproblem \eqref{Eq:maxtotalmn} is reduced into the problem as follows:
\begin{subequations} \label{Eq:maxtotalxi}
\begin{align}
\max_{\xi_{m,n}} &\frac{R_{m,n}(\sum\limits_{k = 1}^{K_{m,n}^{\rm Cu}} I_{m,k} ~\eps_{m,n}(\xi_{m,n})+\!\!\sum\limits_{k = K_{m,n}^{\rm Cu}+1}^{K_{m}}\!\!I_{m,k})}{\xi_{m,n}\sum\limits_{k = 1}^{K_{m,n}^{\rm Cu}} {d_{m,k}^{\rm T}} + \!\!\sum\limits_{k = K_{m,n}^{\rm Cu}+1}^{K_{m}} {d_{m,k}^{\rm T}}}   \\
\text{s.t.} ~& t_{m,n}^{\rm S}\!(\xi_{m,n})\!\!+\!t_{m,n}^{\rm R}\!(\xi_{m,n})\!\!\le\!t_m^{\max}\!\!\!-\!t_{m,n}^{\rm U}\!\!\!-\!t_{m,n}^{\rm T}\!\!\!-\!t_{m,n}^{\rm E}, \label{Eq:delay1} \\
&\eps_{m,n}(\xi_{m,n}) \ge \eps_m^{\rm th},  \label{Eq:xii} \\
&\xi_{m,n} \in [0,1]. \label{Eq:xiii}
\end{align}
\end{subequations}
As ${\cal K}_{m,n}^{\rm Cu}$ is given, the second term in the bracket in the numerator and the second term in the denominator in the objective function are both constants; all the times only related to variable ${\cal K}_{m,n}^{\rm Cu}$ in constraint \eqref{Eq:delay} becomes constants, so that the completion time constraint can be rewritten as \eqref{Eq:delay1}; constraints \eqref{Eq:xi} and \eqref{Eq:C4} are still the same; constraint \eqref{Eq:C5} is no longer needed.
Problem \eqref{Eq:maxtotalxi} is a nonlinear non-convex optimization problem. Fortunately, the semantic accuracy function in the objective is a monotonically increasing function of $\xi_{m,n}$, which can be proved by the first-order derivative of the function. Thus, problem \eqref{Eq:maxtotalxi} can be transformed into a monotonic optimization problem and solved using Polyblock outer approximation algorithm \cite{zhang2013monotonic,Ghanem2020Resource} optimally. Note that problem \eqref{Eq:maxtotalmnE} is the same type as problem \eqref{Eq:maxtotalxi}, so that it can be solved by the same algorithm.

{\it Problem transformation}: For simplicity, we rewrite the numerator of objective function by defining $h(\xi_{m,n})= R_{m,n}(\sum_{k = 1}^{K_{m,n}^{\rm Cu}} I_{m,k}\eps_{m,n}(\xi_{m,n})+\sum_{k = K_{m,n}^{\rm Cu}+1}^{K_{m}}I_{m,k})$. Since $\eps_{m,n}(\xi_{m,n})$ is a monotonically increasing function of $\xi_{m,n}$ and other terms are constants, $h(\xi_{m,n})$ is also a monotonically increasing function of $\xi_{m,n}$. The denominator of objective function is defined as $a_{m,n}\xi_{m,n}+b_{m,n}$, where $a_{m,n}=\sum_{k = 1}^{K_{m,n}^{\rm Cu}} {d_{m,k}^{\rm T}}$ and $b_{m,n}=\sum_{k = K_{m,n}^{\rm Cu}+1}^{K_{m}} {d_{m,k}^{\rm T}}$ are both non-negative constants. Therefore, the objective function is rewritten as $\gamma_{m,n}=\frac{h(\xi_{m,n})}{a_{m,n}\xi_{m,n}+b_{m,n}}$.
Without loss of generality, we apply the logarithmic transformation on the objective function, resulting in $\max_{\xi_{m,n}} {\ln(h(\xi_{m,n}))-\ln(a_{m,n}\xi_{m,n}+b_{m,n})}$. Note that monotonic optimization problems require that the objective functions should be monotonically increasing functions. By introducing an auxiliary variable $\eta$, problem \eqref{Eq:maxtotalxi} can be transformed equivalently into a monotonic optimization problem as follows:
\begin{subequations} \label{Eq:maxMO}
\begin{align}
&~~~~\max_{\xi_{m,n}, \eta} \ln(h(\xi_{m,n}))+\eta-\ln(a_{m,n}\xi_{m,n}^{\max}+b_{m,n})   \\
&\text{s.t.} \ln(a_{m,n}\xi_{m,n}^{\max}\!+\!b_{m,n})\!-\!\eta \ge \ln(a_{m,n}\xi_{m,n}\!+\!b_{m,n}), \\
&~~~~t_{m,n}^{\rm S}\!(\xi_{m,n})\!\!+\!t_{m,n}^{\rm R}\!(\xi_{m,n})\!\!\le\!t_m^{\max}\!\!-\!t_{m,n}^{\rm U}\!\!-\!t_{m,n}^{\rm T}\!\!-\!t_{m,n}^{\rm E}, \label{Eq:delay2} \\
&~~~~\xi_{m,n}^{\rm th}\le \xi_{m,n} \le 1,   \label{Eq:xi2} \\
&~~~~\eta \ge 0,
\end{align}
\end{subequations}
where the objective function is a monotonically increasing function of variables $\xi_{m,n}, \eta$; constraint \eqref{Eq:xi2} is obtained from constraints \eqref{Eq:xii} and \eqref{Eq:xiii}. Specifically, since $\eps_{m,n}(\xi_{m,n})$ is a monotonically increasing function, it is easy to obtain $\xi_{m,n} \ge \xi_{m,n}^{\rm th}$ from \eqref{Eq:xii}, where $\xi_{m,n}^{\rm th}$ is the minimum semantic extraction ratio satisfying the minimum semantic accuracy constraint. Note that $\xi_{m,n}^{\max}$ is the maximum possible value of $\xi_{m,n}$, which could be one. Thus, problem \eqref{Eq:maxMO} can be simplified, i.e.,
\begin{subequations} \label{Eq:maxMO1}
\begin{align}
&\max_{\xi_{m,n}, \eta} \ln(h(\xi_{m,n}))+\eta-\ln(a_{m,n}+b_{m,n})   \\
\text{s.t.} &~ 0 \le \eta \le \ln(a_{m,n}\!+\!b_{m,n})-\ln(a_{m,n}\xi_{m,n}\!+\!b_{m,n}), \label{Eq:eta}\\
&~t_{m,n}^{\rm S}\!(\xi_{m,n})\!+\!t_{m,n}^{\rm R}\!(\xi_{m,n})\!\le\!t_m^{\max}\!\!-\!t_{m,n}^{\rm U}\!\!-\!t_{m,n}^{\rm T}\!\!-\!t_{m,n}^{\rm E}, \label{Eq:delay3} \\
&~\xi_{m,n}^{\rm th}\le \xi_{m,n} \le 1.   \label{Eq:xi3}
%
\end{align}
\end{subequations}
Monotonic optimization problem formulated in \eqref{Eq:maxMO1} can be solved efficiently by Polyblock outer approximation algorithm.

\begin{figure*}[t]
  \centering
  \includegraphics[width=0.79\textwidth]{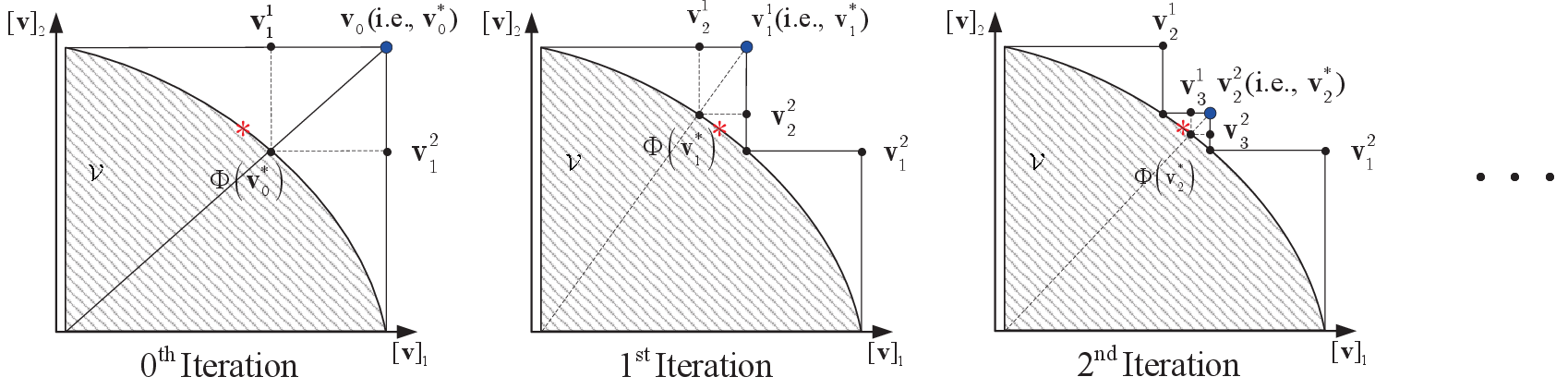}
  \caption{Procedure of polyblock outer approximation algorithm where ${\bf v}$ is a two-dimension vector, the red star is the optimum solution and the blue spot is the current optimum polyblock vertex in each iteration.}
  \label{fig:3}
\end{figure*}

\begin{algorithm}[t]
\caption{Polyblock outer approximation algorithm} \label{algo:2}
\small{
\begin{algorithmic}[1]
\State Initialize $i=0$, polyblock ${\cal P}_0$ as box $[{\bf v}^{\min}, {{\bf v}^{\max}}]$ with vertex set ${\cal V}_0=\{{\bf v}^{\max}\}$, $UB_0=\Gamma({\bf v}_0^*)$, $LB_0=-\infty$, $o_1\ll 1$
\While {$UB_i > (1+o_1) LB_i$}
\State $i=i+1$
\State Find optimum polyblock vertex ${\bf v}_{i-1}^*$ in ${\cal V}_{i-1}$ by \eqref{vstar}
\State Solve projection problem \eqref{Eq:delta} to obtain $\Phi({\bf v}_{i-1}^*)$ and $\delta^*$ using Algorithm \ref{algo:3}
\State Generate new vertices ${\bf v}_i^l$ $\forall l$ by \eqref{newv} and obtain updated vertex set ${\cal V}_i$ by \eqref{Vi}
\State Construct smaller polyblock ${\cal P}_i$ with vertex set ${\cal V}_i$
\State {Update upper bound $UB_i = \Gamma({\bf v}_i^*)$ and lower bound $LB_i = \max\{\Gamma(\Phi({\bf v}_{i-1}^*)), LB_{i-1}\}$}
\EndWhile
\State\Return ${\bf v}^*={\bf v}_i^*$
\end{algorithmic}
}
\end{algorithm}

{\it Polyblock outer approximation:} We first introduce some mathematical preliminaries used in the algorithm.
\begin{itemize}
  \item Box: Given any vector ${\bf{v}} \in \mathbb{R}_{+}^L$, the hyper rectangle $[{\bf v}^{\min}, {{\bf v}^{\max}}]=\{{\bf v}| {\bf v}^{\min} \le {\bf v}\le {{\bf v}^{\max}}\}$ is referred to as a box with vertices ${\bf v}^{\min}$ and ${{\bf v}^{\max}}$.
  \item Polyplock: Given any finite set ${\cal V} \subseteq \mathbb{R}_{+}^L$, the union of all boxes $[{\bf v}^{\min}, {\bf v}]$, ${\bf v} \in {\cal V}$, is a polyblock with vertex set ${\cal V}$.
  \item Normal set: A set ${\cal V} \subseteq \mathbb{R}_{+}^L$ is normal if the box $[{\bf v}^{\min}, {\bf v}] \subseteq {\cal V}$ for any element ${\bf v} \in {\cal V}$.
  \item Projection: For any normal non-empty closed set ${\cal V} \subseteq \mathbb{R}_{+}^L$ and any vertex ${\bf v}$, the projection of ${\bf v}$ on the boundary of ${\cal V}$ is denoted by $\Phi({\bf v})={\bf v}^{\min}+\delta^*({\bf v}-{\bf v}^{\min})$, where $\delta^*=\max\{\delta|{\bf v}^{\min}+\delta({\bf v}-{\bf v}^{\min}) \in {\cal V}, 0 \le \delta \le 1\}$.
\end{itemize}

Problem \eqref{Eq:maxMO1} belongs to the family of monotonic optimization problems. For simplicity, it can be represented in the following standard form:
\begin{align} \label{Eq:maxMO2}
\max_{\bf v} &~~\Gamma({\bf v})   \\
\text{s.t.} &~  {\cal V}=\{{\bf v}| \eqref{Eq:eta}-\eqref{Eq:xi3} \}.     \nonumber
\end{align}
where ${\bf v}=[\xi_{m,n}, \eta]$ is a two-dimension vector. By analyzing constraints, it can be seen that feasible set ${\cal V}$ is a subset of box $[{\bf v}^{\min}, {{\bf v}^{\max}}]$, where ${\bf v}^{\min}=[\xi_{m,n}^{\rm th}, 0]$ and ${{\bf v}^{\max}}=[1, \ln(a_{m,n}\!+\!b_{m,n})-\ln(a_{m,n}\xi_{m,n}^{\rm th}\!+\!b_{m,n})]$.
Since objective function $\Gamma({\bf v})$ is monotonically increasing with respect to ${\bf v}$, optimum solution ${\bf v}^*$ must lie on the boundary of feasible region ${\cal V}$. Hence, optimum solution ${\bf v}^*$ on the boundary can be ultimately approached by utilizing multiple polyblocks iteratively.
We give the detailed process for finding the optimum solution in a general case with $L$-dimension vector ${\bf v}$, which is summarized into the following steps and in Algorithm \ref{algo:2}:

{\bf Step 1}: We establish an initial polyblock as box $[{\bf v}^{\min}, {{\bf v}^{\max}}]$, that contains the whole feasible set ${\cal V}$. We define the polyblock as ${\cal P}_0$ with vertex set ${\cal V}_0=\{ {\bf v}_0\}$, where ${\bf v}_0= {\bf v}^{\max}$.

{\bf Step 2}: Let the vertex of a polyblock that achieves the maximum objective value be the current optimum polyblock vertex. In $i$-th iteration, we find the optimum polyblock vertex in ${\cal V}_{i-1}$, i.e.,
\begin{equation}\label{vstar}
  {\bf v}_{i-1}^*=\arg\max_{{\bf v} \in {\cal V}_{i-1}} \Gamma({\bf v}).
\end{equation}

{\bf Step 3}: We project optimum polyblock vertex ${\bf v}_{i-1}^*$ onto the boundary of feasible region ${\cal V}$ to obtain a new vector $\Phi({\bf v}_{i-1}^*)$. The method to perform the projection of a vertex will be given afterwards. Furthermore, $L$ new vertices can be generated from $\Phi({\bf v}_{i-1}^*)$, which can be expressed as
\begin{equation}\label{newv}
  {\bf v}_i^l={\bf v}_{i-1}^*-([{\bf v}_{i-1}^*-\Phi({\bf v}_{i-1}^*)]_l){\bf e}^l,
\end{equation}
for all $1\le l \le L$, where $[\cdot]_l$ represents the $l$-th element of a vector and ${\bf e}^l$ is an unit vector with a value of one in $l$-th element and zeros in all other elements.

{\bf Step 4}: In $i$-th iteration, a smaller polyblock ${\cal P}_i$ with updated vertex set ${\cal V}_i$ can be constructed, where ${\cal V}_i$ is updated by deleting ${\bf v}_{i-1}^*$ from ${\cal V}_{i-1}$ and adding $L$ new vertices generated from \eqref{newv}, i.e.,
\begin{equation}\label{Vi}
  {\cal V}_i=({\cal V}_{i-1}/\{{\bf v}_{i-1}^*\}) \cup \{{\bf v}_i^1, {\bf v}_i^2, \ldots, {\bf v}_i^L\}.
\end{equation}
Meanwhile, we denote the upper and lower bounds of objective function value in $i$-th iteration by $UB_i$ and $LB_i$, respectively. In this case, they can be defined as follows:
\begin{align}\label{LBUB}
  &UB_i \triangleq \Gamma({\bf v}_i^*), \\
  &LB_i \triangleq \max\{\Gamma(\Phi({\bf v}_{i-1}^*)), LB_{i-1}\},
\end{align}
where $LB_0 = -\infty$.

{\bf Step 5}: Repeat steps 2-4, so that with iteration continues, we can construct a sequence of polyblocks outer approximating the feasible set ${\cal V}$:
\begin{equation}\label{multiP}
  {\cal P}_0 \supseteq {\cal P}_1 \supseteq \cdots \supseteq {\cal P}_i \supseteq \cdots \supseteq {\cal V}.
\end{equation}
Hence, the upper bound will iteratively decrease due to the smaller polyblock. Finally, the iteration will terminate when $UB_i \le (1+o_1) LB_i$, where $o_1$ is the error tolerance of desired accuracy level for approximation.

\begin{algorithm}[t]
\caption{Binary search algorithm used in projection} \label{algo:3}
\small{
\begin{algorithmic}[1]
\State Initialize $j=0$, $\delta^{\min}=0$, $\delta^{\max}=1$, and $o_2 \ll 1$
\While {$\delta^{\max}- \delta^{\min}> o_2$}
\State $j=j+1$, $\delta_j=(\delta^{\max}+\delta^{\min})/2$
\State Check if vector ${\bf v}^{\min}+\delta_j({\bf v}_{i-1}^*-{\bf v}^{\min})$ is in feasible set ${\cal V}$
\If {{\it Yes}}
\State $\delta^{\min}=\delta_j$
\Else
\State $\delta^{\max}=\delta_j$
\EndIf
\EndWhile
\State\Return $\delta^*=\delta^{\max}$
\end{algorithmic}
}
\end{algorithm}

Here, we give the detailed method to obtain projection $\Phi({\bf v}_{i-1}^*)$. From the definition of projection, we know
\begin{equation}\label{phi}
  \Phi({\bf v}_{i-1}^*)={\bf v}^{\min}+\delta^*({\bf v}_{i-1}^*-{\bf v}^{\min}),
\end{equation}
where $\delta^* \in [0, 1]$ is the maximum value remaining $\Phi({\bf v}_{i-1}^*)$ in feasible region ${\cal V}$. Thus, the following problem can be formulated:
\begin{subequations} \label{Eq:delta}
\begin{align}
\max_{\delta} &~~\delta   \\
\text{s.t.} &~ {\bf v}^{\min}+\delta({\bf v}_{i-1}^*-{\bf v}^{\min}) \in {\cal V}, \\
&~ \delta \in [0, 1].
\end{align}
\end{subequations}
The objective in problem \eqref{Eq:delta} is continuous, monotonic, and bounded, so that binary search method can be applied to find the optimum $\delta^*$, as shown in Algorithm \ref{algo:3}. In each iteration, we can decide to increase or decrease $\delta$ by checking the feasibility of resulted vector ${\bf v}^{\min}+\delta({\bf v}_{i-1}^*-{\bf v}^{\min})$ in feasible region.

\vspace{-4mm}
\subsection{SBS association subproblem}
\label{subsec:subproblem2}

After obtaining solutions ${\cal K}_{m,n}^{\rm Cu*}$'s and $\xi_{m,n}^*$'s of joint KUER subproblems for all MD $m$ and SBS $n$, we feed them into original optimization problem \eqref{Eq:maxtotal} and obtain the following SBS association subproblem:
\begin{subequations} \label{Eq:maxUA}
\begin{align}
\max_{x_{m,n}}&\sum\limits_{m = 1}^M\sum\limits_{n = 1}^N {x_{m,n}  \gamma_{m,n}({\cal K}_{m,n}^{\rm Cu*}, \xi_{m,n}^*)   }   \\
\text{s.t.}
%
%
&~\eqref{UA}, \eqref{BA}, \eqref{Eq:C3}.
\end{align}
\end{subequations}
Subproblem \eqref{Eq:maxUA} is a constrained binary linear programming problem. Through observation, it can be mapped into an optimum matching problem in a weighted bipartite graph and solved by modified K-M algorithm optimally and efficiently.

\vspace{-2mm}
\section{Simulation Results}
\label{sec:simulation}

\begin{table}[!t]
\begin{center}
\caption{Default Parameter Settings}
\label{parameters}
\scriptsize{
\begin{tabular}{|c|c|c|c|}
\hline
Parameters     &     Values  & Parameters     &     Values\\
\hline
$I_{m,k}$ &  $U[2,20]$ M suts/s & $f_n^{\rm C}$   &  2 G Hz\\
\hline
$d_{m,k}^{\rm K}$ &  $U[5,80]$ M bits & $p_m^{\rm{T}}$ &  0.1 W\\
\hline
$d_{m,k}^{\rm T}$ &  $U[20,100]$ M bits & $W$ & 10 M Hz \\
\hline
$c_{m,k}$ &  $U[1,100]$ M CPU cycles & $\sigma^2$ &  -120 dBm\\
\hline
$\eps^{\rm th}_{m}$ &  $U[0.7,0.85]$ & $S_n^{\max}$ & 2 \\
\hline
$t^{\max}_{m}$ &  $U[4500,5500]$ ms & $\rho$   &  1  \\
\hline
\end{tabular}
}
\end{center}
\end{table}

In this section, simulation results are presented to demonstrate the superior performance of proposed algorithms. We will compare the proposed efficient algorithm with proposed optimum algorithm and another no knowledge sharing scheme in \cite{KBC}.
That is, semantic communications is performed based on initial matched KBs without knowledge sharing and the solutions can be obtained optimally by our proposed algorithm.
Both distance-based path loss and small-scale fading are considered in link gains, given as $g_{m,n}=10^{-3}{\rho_{m,n}}^2 d_{m,n}^{-2}$, where $d_{m,n}$ is the distance between MD $m$ and SBS $n$ and ${\rho_{m,n}}^2$ is a random variable with exponential distribution and unit mean, since $\rho_{m,n}$ is the additional Rayleigh distributed small-scale channel fading \cite{DT,TWC}. Note that a 30 dB average signal power attenuation is assumed at a reference distance of 1 meter.
We adopt the semantic accuracy nonlinear model with $(\theta_1,\theta_2,\theta_3,\theta_4)=(-6.205e-8,16.45,0.9228,-0.06917)$ used in \cite{Liu2024Adaptable} to estimate the relationship between the semantic accuracy and extraction ratio.
Table~\ref{parameters} summarizes the default parameters, where the parameter values refer to \cite{Liu2024Adaptable,Chen2024Semantic} and $U[a,b]$ represents the uniform distribution between $a$ and $b$.
The numerical results below are based on two different scenarios. The first scenario is to compare the performance of proposed efficient algorithm with optimum algorithm and no knowledge sharing scheme. Due to the high complexity of running the optimum algorithm when the number of variables is large, small size networks (with a small number of SBSs, MDs, and required knowledge classes) are considered in this scenario. The small gap between the efficient algorithm and optimum solution shows the excellent performance of the former. In the second scenario, networks with relatively large number of SBSs, MDs, and knowledge classes are simulated to compare the performance of proposed efficient algorithm with no knowledge sharing scheme, and these results help demonstrate how much the joint knowledge sharing, SBS association, and semantic extraction ratio in proposed algorithm contributes to the superior performance of task-oriented hybrid communications.

\vspace{-3mm}
\subsection{Scenario 1: small size networks}
\label{subsec:scenario1}

We first compare the optimum results from proposed optimum algorithm with the proposed efficient algorithm and no knowledge sharing scheme. Due to the high complexity of optimum algorithm, a small size network is simulated. We consider a network of a circular area with a radius of 150 m centered at the origin. The network consists of 3 SBSs and 5 MDs. The three SBSs are located at (0, 75 m), (-75 m, -75 m), and (75 m, -75 m), and the locations of all MDs are uniformly distributed in the service area. There are 10 knowledge classes in the system, where the KB at each SBS stores 6 knowledge classes randomly picked from 10 classes and each MD requires 3 to 6 knowledge classes to perform semantic communications. The simulation results are obtained by averaging over 100 independent experiments, each of which is based on one set of randomly generated MD locations, and task and knowledge parameters.

\begin{figure}[t]
  \centering
  \includegraphics[height=65mm,width=75mm]{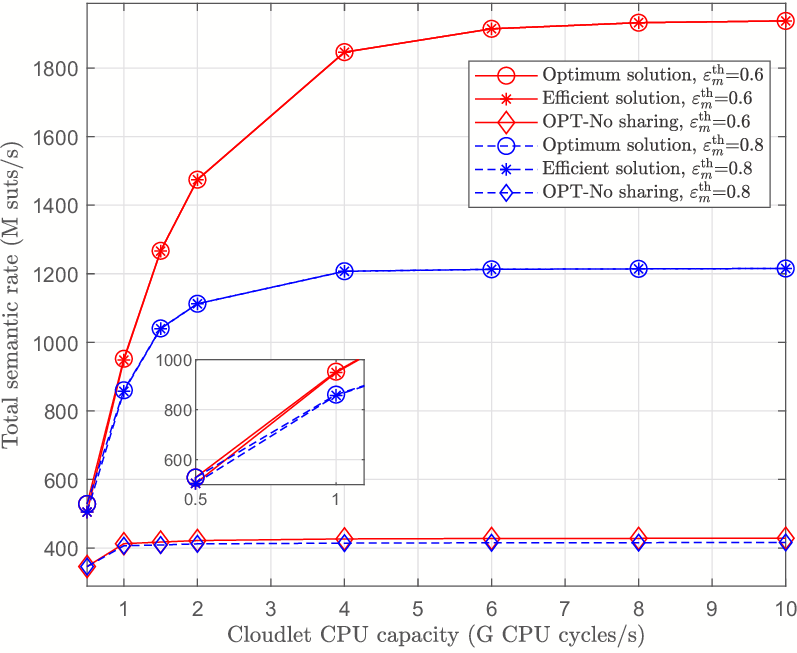}
  \caption{Total semantic rate versus cloudlet computation capacity.}
  \label{fig:4}
  \vspace{-3mm}
\end{figure}

Fig.~\ref{fig:4} shows the total semantic transmission rate of all MDs versus the cloudlet computation capacity $f_n^{\rm C}$ (same for all cloudlets). It shows that the achieved total semantic rate increases significantly with the increase of cloudlet computation capacity when $f_n^{\rm C}$ is relatively small and then becomes a constant. It is because when $f_n^{\rm C}$ is large enough, the performance is no longer constrained by cloudlet computation capacity but affected by the other factors, such as wireless transmission conditions and semantic accuracy requirements.
It can also be seen that the obtained efficient solutions are close to the optimum solutions. It verifies the effectiveness of our proposed efficient solution. The gap between the optimum and efficient solutions decreases with the increase of cloudlet computation capacity, since there is more time left for wireless transmissions when $f_n^{\rm C}$ is relatively large, more MDs can upload all the mismatched knowledge to perform semantic communications. The solutions obtained from both algorithms for knowledge sharing are same, i.e., uploading all the mismatched knowledge. In addition, we can see that the total semantic rate of both algorithms is much higher than the comparison, i.e., the optimum solution without knowledge sharing. It demonstrates the significant importance of knowledge sharing in semantic communication performance enhancement, especially when the network resources are limited. We can also see that the total semantic rate obtained from all the solutions increases with the decrease of semantic accuracy requirements of MDs, since MDs can extract and transmit less data from raw data with lower semantic accuracy requirements to increase semantic transmission rate.

\begin{figure}[t]
  \centering
  \includegraphics[height=65mm,width=75mm]{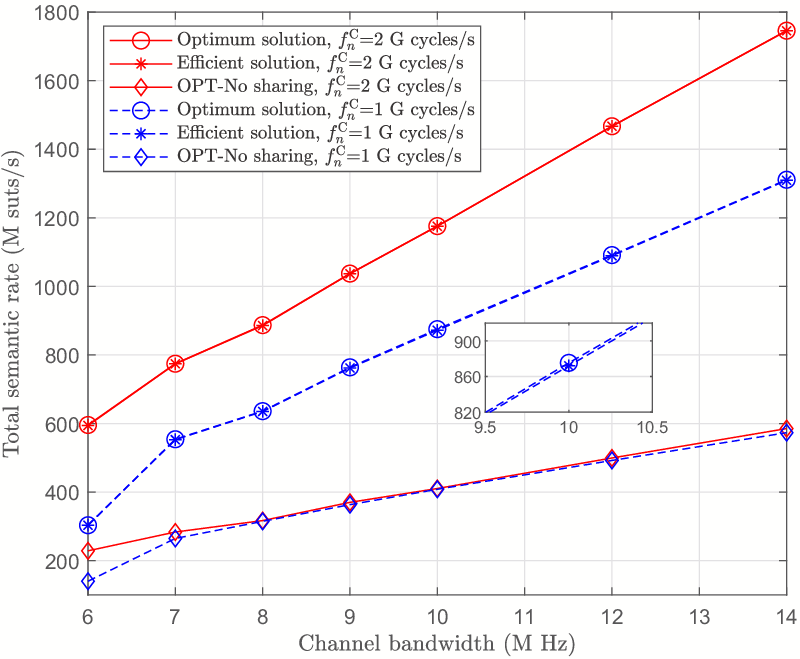}
  \caption{Total semantic rate versus wireless channel bandwidth.}
  \label{fig:5}
\vspace{-5mm}
\end{figure}

\begin{figure}[t]
  \centering
  \includegraphics[height=65mm,width=75mm]{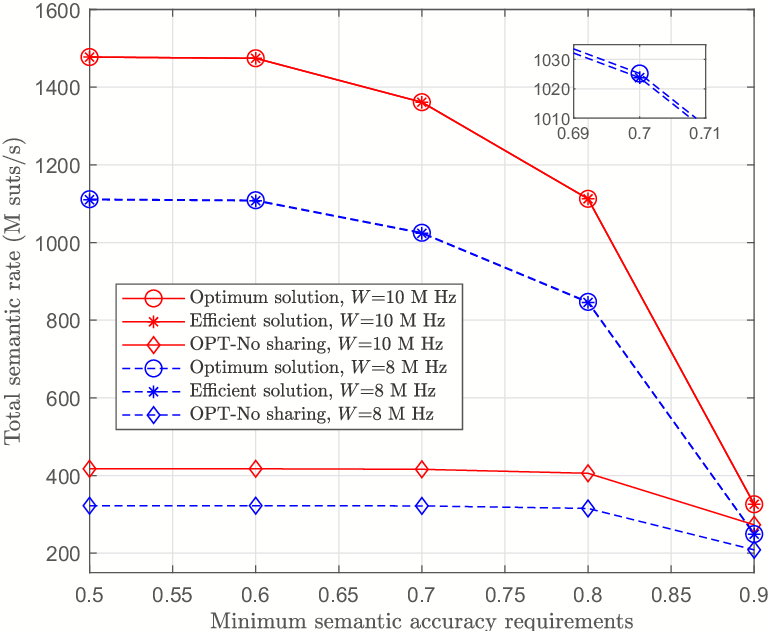}
  \caption{Total semantic rate versus minimum semantic accuracy requirements.}
  \label{fig:6}
  \vspace{-5mm}
\end{figure}

Fig.~\ref{fig:5} shows the total semantic transmission rate of all MDs versus the wireless channel bandwidth $W$. The total semantic rate of all MDs obtained from all solutions increases with the increase of channel bandwidth. When $W$ is relatively large, the total semantic rate is proportional to wireless channel bandwidth, since the total semantic rate is proportional to the wireless transmission rate observed by the definition of semantic rate while the wireless uploading time decreases with the increase of channel bandwidth, the task deadlines can be satisfied with high probability. It can be seen that the total semantic rate obtained from efficient solution is extremely close to that obtained from the optimum solution, and is much higher than that obtained from the optimum solution without knowledge sharing. It further demonstrates the superior performance of our proposed communication mechanisms and solutions. It also shows that the total semantic rate when $f_n^{\rm C}=2$ G cycles/s is higher than that when $f_n^{\rm C}=1$ G cycles/s, which is consistent with Fig.~\ref{fig:4}.

Fig.~\ref{fig:6} shows the total semantic transmission rate of all MDs versus the minimum semantic accuracy requirements $\eps_m^{\rm th}$ (same for all MDs). The total semantic transmission rate decreases slowly when $\eps_m^{\rm th}$ is relatively small and then drops dramatically when $\eps_m^{\rm th}$ is sufficiently large. The total semantic rate decreases with the increase of minimum semantic accuracy requirements of MDs, since semantic information data must be extracted more from raw data in order to satisfy higher semantic accuracy requirement, resulting in lower semantic transmission rate. When $\eps_m^{\rm th}$ is relatively small, the semantic performance is mainly decided by the delay constraints; when $\eps_m^{\rm th}$ becomes larger, the achieved semantic accuracy matters a lot. We can also see that the total semantic rate obtained from the proposed mechanisms and solutions, jointly considering knowledge sharing, SBS association, and semantic extraction ratio, is more sensitive to the change of minimum semantic accuracy requirements than that obtained from the optimum solution without considering knowledge sharing. Nevertheless, the proposed solutions have superior performance compared to the comparison. The other observations are similar to Fig.~\ref{fig:5}.

\vspace{-3mm}
\subsection{Scenario 2: large size networks}
\label{subsec:scenario2}

We further compare the proposed efficient solution with no knowledge sharing scheme in \cite{KBC} by simulating relatively larger size networks. We consider a network of a circular area with a radius of 300 m centered at the origin. The network consists of 5 SBSs and 10 MDs. The 5 SBSs are located at (0, 0), (150 m, 0), (0, 150 m), (-150 m, 0), and (0, -150 m), and the locations of all MDs are uniformly distributed in the service area. There are 20 knowledge classes in the system, where the KB at each SBS stores 8 knowledge classes randomly picked from 20 classes and each MD requires 11 to 15 knowledge classes to perform semantic communications.

\begin{figure}[t]
  \centering
  \includegraphics[height=65mm,width=75mm]{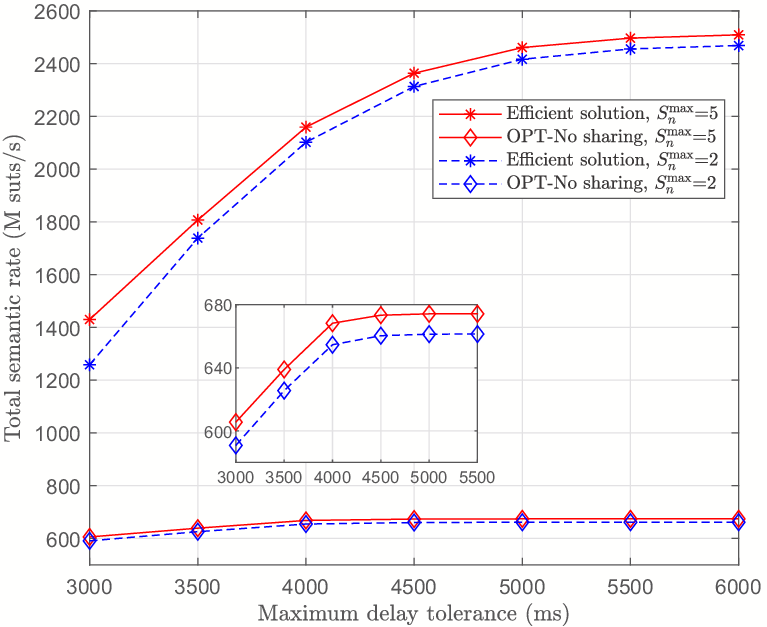}
  \caption{Total semantic rate versus maximum delay tolerance.}
  \label{fig:7}
\vspace{-5mm}
\end{figure}

Fig.~\ref{fig:7} shows the total semantic transmission rate of MDs versus the maximum delay tolerance $t_m^{\max}$ (same for all MDs). As the maximum delay tolerance of MDs increases, the total semantic rate of all solutions increases significantly when $t_m^{\max}$ is relatively small and then becomes a constant. When the maximum delay tolerance is relatively tight, the increase of it can allow more mismatched knowledge data to be uploaded and shared to the KB of associated SBS, leading to higher effective semantic transmission rate; as $t_m^{\max}$ is sufficiently large, the task delay tolerance can always be satisfied so that the network performance is no longer affected by task delay constraints. It also shows that the total semantic rate of efficient solutions is much higher than that of no knowledge sharing scheme in this relatively larger size network, which demonstrates the efficiency and effectiveness of the proposed solution and the benefits of jointly considering knowledge sharing in our proposed mechanism. We can see that the total semantic rate when $S_n^{\max}=5$ is higher than that when $S_n^{\max}=2$, since there are more cloudlets at the SBSs so that MDs can have chance to access the cloudlet with better channel conditions.

\begin{figure}[t]
  \centering
  \includegraphics[height=65mm,width=75mm]{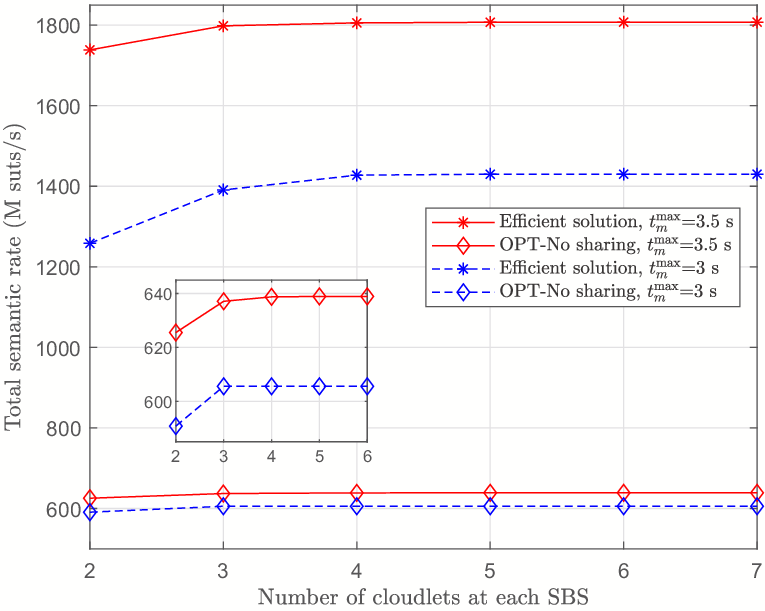}
  \caption{Total semantic rate versus number of cloudlets at each SBS.}
  \label{fig:8}
\end{figure}

Fig.~\ref{fig:8} shows the total semantic transmission rate of MDs versus the number of cloudlets at each SBS $S_n^{\max}$ (same for all SBSs). The total semantic rate of all MDs obtained from our proposed efficient solution is much higher than that obtained from the optimum solution without knowledge sharing. The total semantic rate of all MDs when $t_m^{\max}=3.5$ s is higher than that when $t_m^{\max}=3$ s in both solutions. With the increase of $S_n^{\max}$, the total semantic rate of MDs for all solutions increases when $S_n^{\max}$ is relatively small. The increase becomes saturated when the number of cloudlets at each SBS is sufficiently large. It is because MDs can access the cloudlets at SBSs having better channel conditions, resulting in higher performance. However, when the number of cloudlets at each SBS is large enough, all the MDs can access the best cloudlet to complete the target tasks so that the SBS associations with the maximum performance in this case can always be obtained.

\begin{figure}[t]
  \centering
  \includegraphics[height=65mm,width=75mm]{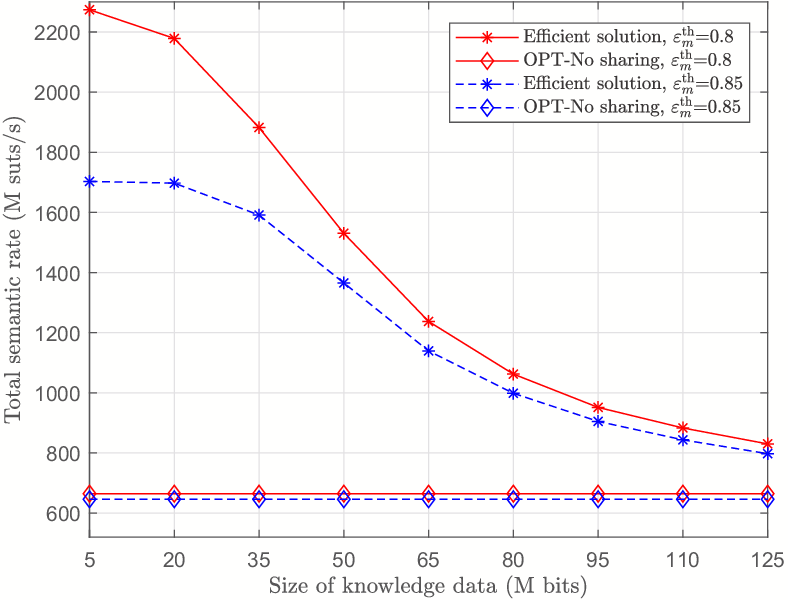}
  \caption{Total semantic rate versus knowledge data size.}
  \label{fig:9}
  \vspace{-5mm}
\end{figure}

Fig.~\ref{fig:9} shows the total semantic transmission rate of MDs versus the knowledge data size $d_{m,k}^{\rm K}$ (same for all MDs and classes). It can be seen that the total semantic rate obtained from no knowledge sharing scheme is a constant as the increase of knowledge data size, since there is no knowledge sharing from MDs to SBSs, the size of knowledge data will not affect the performance. Although the total semantic rate of MDs obtained from proposed efficient solution decreases with the increase of knowledge data size, it is higher than that obtained from the optimum solution without knowledge sharing. The performance gap is significantly large when $d_{m,k}^{\rm K}$ is relatively small and becomes smaller as the increase of $d_{m,k}^{\rm K}$. This indicates that when the knowledge data size is sufficiently large, knowledge sharing in semantic communications loses the benefits due to longer data transmission time. Therefore, when the knowledge data size is below a threshold, the knowledge sharing in semantic communications consumes shorter transmission time and boosts the semantic transmission rate of MDs compared to no knowledge sharing scheme.

\vspace{-4mm}
\section{Conclusions}
\label{sec:conclusions}

In this paper, we proposed a novel knowledge sharing-enabled task-oriented hybrid semantic and bit transmission mechanism to tackle the KB mismatch problem, where a MD can proactively share the task-related mismatched knowledge to the associated SBS and the bit communications are adopted as an aid to transmit the rest data related to unshared mismatched knowledge. In a multi-cell network, the derived generalized effective semantic transmission rate of all MDs was maximized by jointly optimizing knowledge sharing decision, semantic extraction ratio, and SBS association, considering the heterogeneous transceivers, task demands, and channel conditions. The originally formulated MINLP problem was decomposed into multiple subproblems. An optimum algorithm was proposed and another efficient algorithm was further developed using hierarchical class partitioning and monotonic optimization. Simulation results demonstrated the validity and superior performance of proposed solutions. A variety of results were presented that characterize the tradeoffs between task delay tolerance, communication and computation resources, and knowledge data size in terms of total semantic rate. The proposed efficient solution achieved close-to-optimum performance and outperformed the comparisons for a wide range of system parameters.

\vspace{-2mm}
\bibliographystyle{IEEEtran}
\bibliography{IEEEabrv,mybibfileKB}

\end{document}